\documentclass[%
reprint,
 amsmath,amssymb,
 prl,
floatfix,
]{revtex4-2}

\usepackage{graphicx}
\usepackage{dcolumn}
\usepackage{bm}
\usepackage{hyperref}

\begin{document}

\preprint{APS/123-QED}

\title{Fast Ion Isotropization by Current Sheet Scattering in Magnetic Reconnection Jets}

\author{Louis Richard}
 \email{louis.richard@irfu.se}
\affiliation{%
Swedish Institute of Space Physics, Uppsala, Sweden}
\affiliation{Department of Physics and Astronomy, Space and Plasma Physics, Uppsala University, Sweden 
}%

\author{Yuri V. Khotyaintsev}
\author{Daniel B. Graham}
\affiliation{
Swedish Institute of Space Physics, Uppsala, Sweden
}%

\author{Andris Vaivads}
\affiliation{Space and Plasma Physics, School of Electrical Engineering, KTH Royal Institute of Technology, Stockholm, Sweden}

\author{Daniel J. Gershman}
\affiliation{
NASA Goddard Space Flight Center, Greenbelt, Maryland 20771, USA
}%

\author{Christopher T. Russell}
\affiliation{
University of California, Los Angeles, California 90095, USA
}%

\date{\today}

\begin{abstract}
We present a statistical analysis of ion distributions in magnetic reconnection jets using data from the Magnetospheric Multiscale spacecraft. Compared with the quiet plasma in which the jet propagates, we often find anisotropic and non-Maxwellian ion distributions in the plasma jets. We observe magnetic field fluctuations associated with unstable ion distributions, but the wave amplitudes are not large enough to scatter ions during the observed travel time of the jet. We estimate that the phase-space diffusion due to chaotic and quasi-adiabatic ion motion in the current sheet is sufficiently fast to be the primary process leading to isotropization.
\end{abstract}

\maketitle

Fast plasma flows, referred to as jets, are ubiquitous phenomena in the universe \cite{masuda_loop-top_1994,phan_extended_2000,pudritz_magnetic_2012}. The dynamical evolution of the plasma jets is thought to be driven by changes in the magnetic field configuration (e.g., the Blandford-Znajek mechanism \cite{parfrey_first-principles_2019}). Magnetic reconnection, in particular, is a topological reconfiguration of the magnetic field, which results in the conversion of the magnetic field energy into particle energy through particle acceleration and heating \cite{biskamp_magnetic_2000,yamada_magnetic_2010}. Complex non-Maxwellian ion velocity distribution functions (VDFs) characterize collisionless plasmas \cite{hoshino_ion_1998}. For example, cold Alfv\'enic counter-streaming ion beams entering into the reconnection exhaust \cite{nagai_counterstreaming_2002,eastwood_ion_2015} provide a source of free energy for ion beam-driven instabilities \cite{liu_ionbeamdriven_2019}. Simulations \cite{drake_ion_2009,hietala_ion_2015} have shown that as the beams cross the reconnection jet, heating of the beams results in the formation of a single VDF with temperature anisotropy $R_i = T_{i\perp} / T_{i\parallel} < 1$, where $T_{i\parallel}$ and $T_{i\perp}$ are the ion temperature parallel and perpendicular to the local magnetic field $\mathbf{B}$. Deviations from the thermodynamic equilibrium can drive micro-instabilities \cite{gary_theory_1993}, contributing to ion heating in the reconnection region.

Spacecraft observations suggest wave generation in the reconnection jet due to ion temperature anisotropy-driven instabilities \cite{voros_magnetic_2011,wu_proton_2013}. As the waves grow and saturate, pitch-angle scattering due to wave-particle interaction limits deviations from isotropy ($T_{i\perp}= T_{i\parallel}$). Simulations indicate that the growth rate of the electromagnetic ion temperature anisotropy-driven instabilities is much slower than the generation of the unstable VDFs~\cite{kunz_firehose_2014,hietala_ion_2015}. Thus, wave-particle interactions can provide isotropization of the ion VDFs in the jets only after hundreds of ion gyroperiods, i.e, after the jet has convected hundreds $d_i$ downstream of the reconnection site \cite{hietala_ion_2015}, with $d_i=c/\omega_{pi}$ the ion inertial length. However, observations show a dominance of isotropic ion VDFs already at distances $\sim 50~d_i$ from the reconnection region \cite{wu_proton_2013}. This suggests that other mechanisms relax the temperature anisotropy in the reconnection jets (e.g., current sheet pitch-angle scattering \cite{eastwood_consistency_1972,wagner_particle_1979,tsyganenko_pitch-angle_1982,birmingham_pitch_1984,buchner_regular_1989,ashour-abdalla_chaotic_1990}). Investigating the processes limiting the ion temperature anisotropy is thus crucial to understanding the energy budget and partition in the reconnection jets.

In this Letter, we use data from the Magnetospheric Multiscale (MMS) spacecraft \cite{burch_magnetospheric_2016} in the Earth's magnetotail to study the temperature anisotropy and non-Maxwellianity of ion VDFs in the reconnection jets. We use 516 jets from a database of jets observed by MMS in the central plasma sheet (CPS) of the Earth's magnetotail ($\beta_i \geq 0.5$~\cite{angelopoulos_statistical_1994}, where $\beta_i = n_i k_B T_i / P_{mag}$, $n_i$ is the ion number density, $T_i$ the ion temperature and $P_{mag} = |\mathbf{B}|^2/2\mu_0$ is the magnetic pressure) \cite{richard_are_2022}. CPS plasma jets are thought to result from magnetic reconnection or kinetic ballooning/interchange instability \cite{pritchett_kinetic_2010}. It is reasonable to assume that the plasma jets are primarily reconnection outflows since we observe the jets relatively close (the median distance is $\delta x_c \approx 2.8~R_E$) to the statistical location of the reconnection X-line in the near-Earth magnetotail at $X_{GSM}\approx -25~ R_E$ in geocentric solar magnetospheric (GSM) coordinates \cite{richard_are_2022,nagai_solar_2005}. The magnetic field is measured by the FGM instrument \cite{russell_magnetospheric_2016}, and the ion VDFs are measured by the FPI-DIS instrument \cite{pollock_fast_2016} with corrections removing a background population to account for the penetrating radiations \cite{gershman_systematic_2019}. To increase the counting statistics and thus reduce the measurement uncertainties, we average the measurements across the four spacecraft and over a $450~\textrm{ms}$ (or three ion VDFs) running time window (see Supplemental Material~\cite{supplemental_material}).

\begin{figure}
\includegraphics[width=\linewidth]{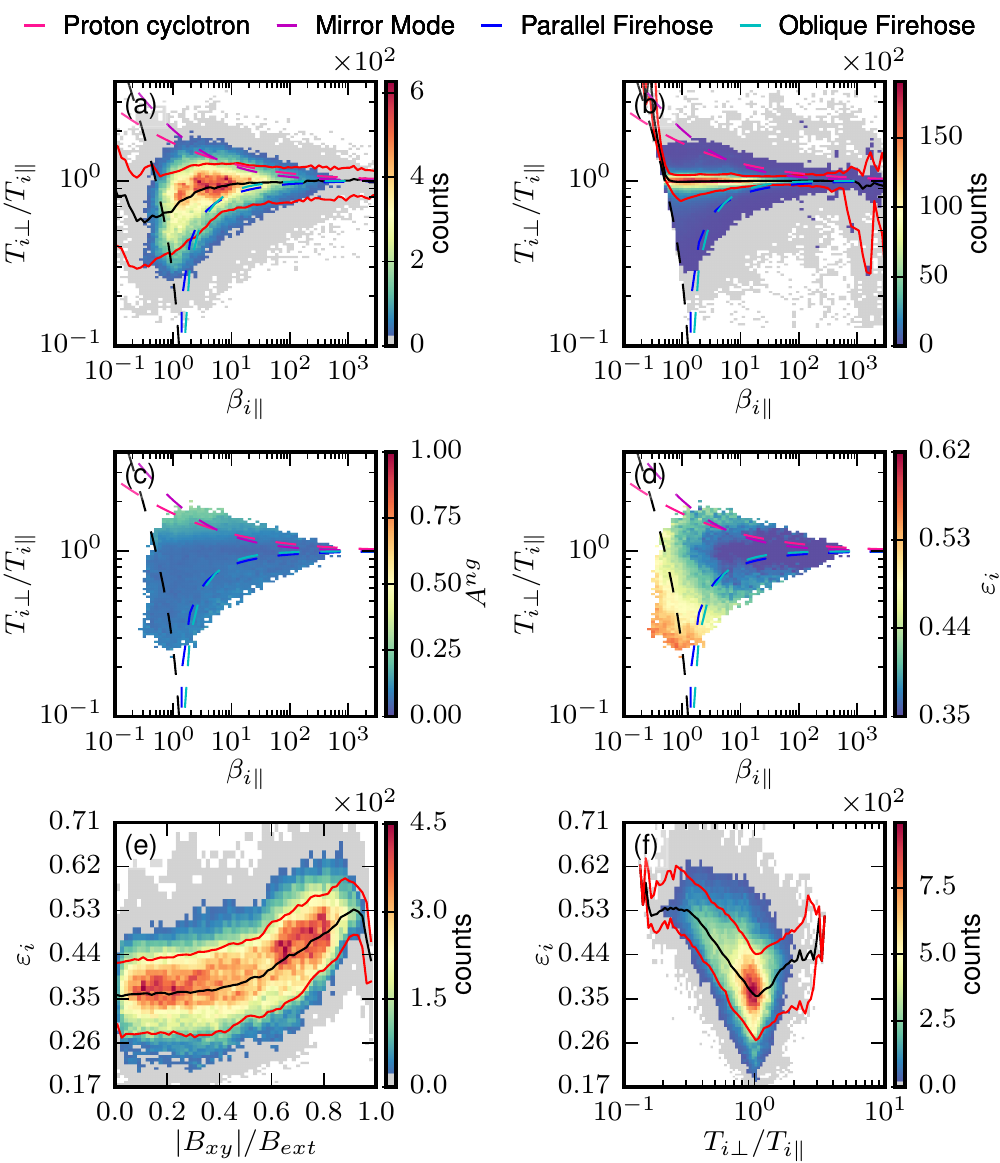}
\caption{\label{fig:distribution}Distribution of measurements. (a) and (b) Histograms of $R_i=T_{i\perp}/T_{i\parallel}$ versus $\beta_{i\parallel}$ in the reconnection jets and quiet CPS ($\beta_i>0.5$, $|V_i| < 100~\textrm{km}~\textrm{s}^{-1}$), respectively. The black line in panels (a)-(b) indicates the median of $R_i$ as a function of $\beta_{i\parallel}$, and the red curves indicate the 10th and 90th percentiles. (c) and (d) Conditional average of non-gyrotropic anisotropy $A^{ng}$ and non-Maxwellianity $\varepsilon_i$ in the $(\beta_{i\parallel}, R_i)$ space. The black dash line in panels (a)-(d) indicates the CPS threshold $\beta_i \geq 0.5$. (e) Histogram of $\varepsilon_i$ versus $|B_{xy}| / B_{ext}$. (f) Histogram of $\varepsilon_i$ versus $R_i$. The black line in panels (e)-(f) indicates the median of $\varepsilon_i$ as a function of $|B_{xy}| / B_{ext}$ and $R_i$ and the red curves indicate the 10th and 90th percentiles. Grey-shaded bins in panels (a), (b), (e), and (f) indicate counts below $5\sigma = 5 \sqrt{n}$.}
\end{figure}

Overall, the ion VDFs are predominantly isotropic $T_{i\perp}\sim T_{i\parallel}$ both in the reconnection jets ($T_{i\perp} / T_{i\parallel}=0.9_{-0.3}^{+0.25}$) [Fig.~\ref{fig:distribution}a] and in the quiet CPS ($T_{i\perp} / T_{i\parallel}=1.0_{-0.13}^{+0.09}$) [Fig.~\ref{fig:distribution}b]. However, compared to the quiet CPS, a significant fraction of the ion VDFs in the jets show a large deviation from $T_{i\perp}= T_{i\parallel}$ (see also Supplemental Material~\cite{supplemental_material}), indicating that magnetic reconnection generates anisotropic ion VDFs.

Here we refer to the anisotropy in the sense of the gyrotropic, double-adiabatic or Chew-Goldberger-Low (CGL) closure~\cite{chew_boltzmann_1956}. Indeed, quantifying the non-gyrotropic anisotropy~\cite{del_sarto_shear-induced_2018} $A^{ng}=(\lambda_2 - \lambda_3)/(\lambda_2 + \lambda_3)$, where $\lambda_i$ are the eigenvalues of the temperature tensor such as $\lambda_1>\lambda_2>\lambda_3$, we observe that the ion VDFs are generally gyrotropic [Fig~\ref{fig:distribution}c]. Hence, in what follows, we use the gyrotropic, double adiabatic anisotropy $T_{i\perp}/T_{i\parallel}$.

Observations in the solar wind \cite{hellinger_solar_2006} have shown that the threshold $\mathcal{R}_i$ for the ion temperature anisotropy driven instabilities scales with $\beta_{i\parallel} = n_i k_B T_{i\parallel} / P_{mag}$ with:

\begin{equation}
    \mathcal{R}_i = 1 + \frac{a}{\left ( \beta_{i\parallel} - \beta_0 \right )^b}\, ,
\end{equation}

\noindent where $a$, $b$, and $\beta_0$ are constants determined from linear Vlasov analysis of the kinetic dispersion relation in a homogeneous, magnetized, collisionless plasma with bi-Maxwellian ion VDFs \cite{gary_theory_1993}. We consider thresholds corresponding to the growth rate $\gamma = 10^{-2}\omega_{ci}$ for a proton-electron plasma with isotropic electrons given in Ref. \cite{verscharen_collisionless_2016}. We observe that $70\%$ of the ion VDFs lie within the instabilities thresholds, which, in analogy to the solar wind~\cite{bale_magnetic_2009}, suggests that the temperature anisotropy is regulated in the reconnection jets by wave-particle interaction. Still, we will show below that this is not the case.

\begin{figure}
\includegraphics[width=\linewidth]{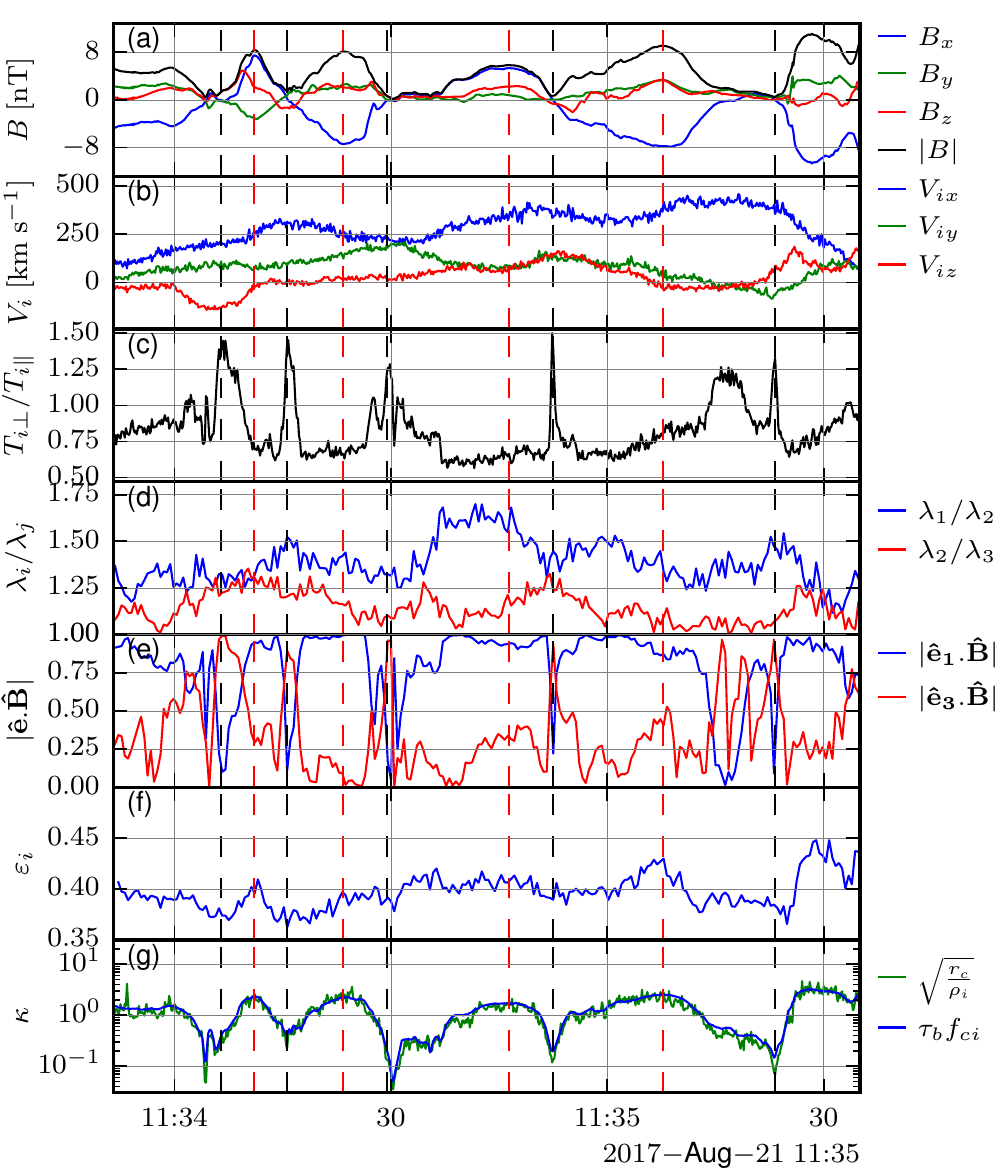}
\caption{\label{fig:example} Example reconnection jet with current sheet flapping motion. (a) Magnetic field. (b) Ion bulk velocity. (c) Ion temperature anisotropy $T_{i\perp}/T_{i\parallel}$, (d) ratios of the eigenvalues of the temperature tensor $\mathbf{T_i}$ with $\lambda_1>\lambda_2>\lambda_3$. (e) $\mathbf{\hat{e}_1}.\mathbf{B}$ and $\mathbf{\hat{e}_3}.\mathbf{B}$ where $\mathbf{\hat{e}_1}$ and $\mathbf{\hat{e}_3}$ are the eigenvectors of $\mathbf{T_i}$ associated with $\lambda_1$ and $\lambda_3$. (f) Non-Maxwellianity $\varepsilon_i$. (g) Adiabaticity parameter $\kappa$ from 4 spacecraft method $\kappa=\sqrt{r_c/\rho_i}$ (green) and time scales ratio $\kappa=\tau_b f_{ci}$ (blue).}
\end{figure}

Fig.~\ref{fig:example} presents an example of a reconnection jet observed by MMS on August 21, 2017. The large amplitude oscillations of the magnetic field indicate the flapping of the magnetotail current sheet (CS) \cite{sergeev_current_2003,wei_observations_2019, richard_observations_2021}. The flapping enables sampling of the ion VDFs across the reconnection jet. We observe $T_{i\perp} < T_{i\parallel}$ at the edges of the CS (local $|\mathbf{B}|$ maxima) and $T_{i\perp}>T_{i\parallel}$ at the CS center (local $|\mathbf{B}|$ minima) [Fig.~\ref{fig:example}c]. Such temperature anisotropies can result from two counter-streaming beams VDFs at the CS edges~\cite{drake_ion_2009,hietala_ion_2015} and Speiser-like meandering ions at the CS center~\cite{artemyev_proton_2010,hietala_ion_2015}. To investigate these large changes in temperature anisotropy, we compute the eigenvalues [Fig.~\ref{fig:example}d] and the eigenvectors of the temperature tensor \cite{del_sarto_shear-induced_2018}. We find that the eigenvectors are nearly constant throughout the interval with $\mathbf{\hat{e}}_1=[-0.99, 0.13, 0.03]$, and $\mathbf{\hat{e}}_3=[-0.01, -0.28,  0.96]$ in GSM coordinates, and it is the rotation of $\mathbf{B}$ with respect to the constant eigenvectors [Fig.~\ref{fig:example}e] causing the large spikes in $T_{i\perp}/T_{i\parallel}$. Such behavior is consistent with the demagnetization of the ions in the CS center. 

Fig.~\ref{fig:example}f shows the ion non-Maxwellianity 

\begin{equation}
    \varepsilon_i = (2n_i)^{-1} \int |f_i - f_{i\mathrm{bM}}| \textrm{d}^3v\, ,
\end{equation}

\noindent which is the normalized ($\varepsilon_i\in[0, 1]$) zeroth-order moment of the difference between the observed ion VDF $f_i$ and a bi-Maxwellian model $f_{i\mathrm{bM}}$ with the same moments \cite{graham_nonmaxwellianity_2021}.
The ion counting statistics are below the instrument noise floor at speeds $v < v_{Ti} / 3$, where $v_{Ti}=\sqrt{2k_B T_i/m_i}$, due to the high ion bulk energy. For this reason, we compute $\varepsilon_i$ using energy channels $K_i > T_i / 9$. $\varepsilon_i$ peaks at the CS edges [Fig.~\ref{fig:example}f]. Simulations have shown that at the exhaust boundaries, the ion VDFs typically consist of cold Alfv\'enic counter-streaming beams~\cite{drake_ion_2009,hietala_ion_2015}. Such VDFs are non-velocity space-filling and thus yield a large $\varepsilon_i$ consistent with our observations. In contrast, $\varepsilon_i$ drops at the center of the CS, consistent with Speiser-like meandering ions filling the velocity space and resulting in a lower $\varepsilon_i$. 

We observe that $\varepsilon_i$ increases with decreasing $\beta_{i\parallel}$ [Fig.~\ref{fig:distribution}d]. We estimate the background value $\varepsilon_i= 0.35\pm 0.09$ in the quiet CPS for comparison (see Supplemental Material~\cite{supplemental_material}). The same analysis repeated for a longer time averaging windows (5, 7, and 9 ion VDFs) gives similar results (see Supplemental Material~\cite{supplemental_material}). Fig.~\ref{fig:distribution}e presents $\varepsilon_i$ as a function $|\mathbf{B}_{xy}| / B_{ext}$, which is a proxy of the distance to the CS center~\cite{bowling_motion_1974,sergeev_current_1998}, and where $B_{ext}=\sqrt{1 + \beta_i}|\mathbf{B}|$ is the external field obtained from pressure balance assumption~\cite{asano_evolution_2003}. We observe that $\varepsilon_i$ increases with $|\mathbf{B}_{xy}| / B_{ext}$  [Fig.~\ref{fig:distribution}e], meaning that deviations from a bi-Maxwellian are statistically more pronounced at the CS edges similar to that seen in Fig.~\ref{fig:example}f. In addition, enhancements of $\varepsilon_i$ are associated with large temperature anisotropy [Fig.~\ref{fig:distribution}d and~\ref{fig:distribution}f], indicating that the magnetic reconnection drives both the temperature anisotropy and non-Maxwellianity.

\begin{figure}
\includegraphics[width=\linewidth]{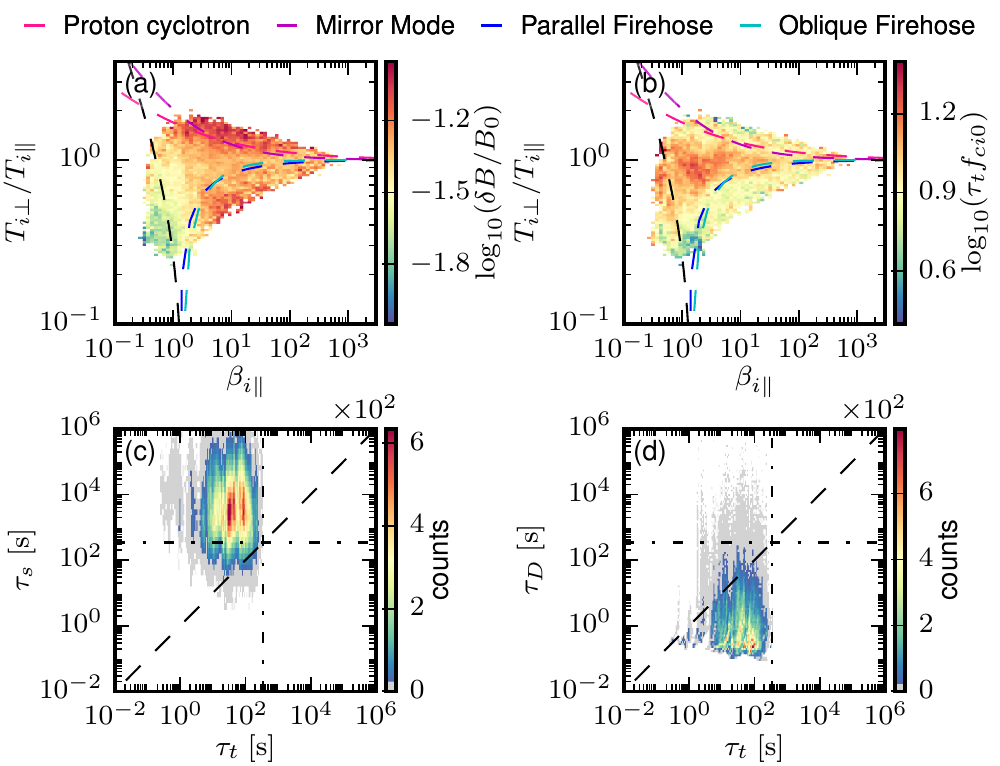}
\caption{\label{fig:time-scales}Comparison of diffusion time scales and travel time of the jet. (a) Magnetic field fluctuation normalized to the background field. (b) The travel time of the jet $\tau_t$. (c) Pitch-angle scattering time $\tau_{s}$ due to Alfv\'enic fluctuations versus $\tau_t$. (d) Phase-space diffusion time $\tau_D$ versus $\tau_t$. Grey-shaded bins in panels (c) and (d) indicate counts below $5\sigma = 5 \sqrt{n}$.}
\end{figure}

The non-equilibrium VDFs can be unstable to a range of instabilities. To investigate this, we compute the magnetic field fluctuations $|\delta \mathbf{B}|/B_0$ normalized to the background magnetic field $B_0=B_{ext}/2$  \cite{artemyev_proton_2010}. To eliminate the large-scale fluctuations, such as flapping, we high-pass filter $\delta \mathbf{B}$ at $f\geq 0.1~\mathrm{Hz}$, which corresponds to the  highest expected flapping frequency~\cite{wei_observations_2019}. Since the energy of the magnetic field fluctuations cascades to smaller scales, the power is dominated by the fluctuations at $f\approx 0.1~\mathrm{Hz}$, i.e., assuming Taylor’s hypothesis applies~\cite{voros_bursty_2006}, $k\rho_{i0}\approx 1.4\pm 0.8$ where $\rho_{i0}=v_{Ti}/\omega_{ci0}$ with $\omega_{ci0}=eB_0/m_i$. We observe strong magnetic wave activity in concert with the unstable ion VDFs [Fig~\ref{fig:time-scales}a] consistent with previous observations \cite{wu_proton_2013,voros_magnetic_2011}. This enhanced wave activity associated with unstable VDFs indicates the growth of the instabilities in the reconnection jets. 

The majority of the observed ion VDFs are nearly isotropic $T_{i\perp}\sim T_{i\parallel}$ [Fig.~\ref{fig:distribution}a], meaning that there must be a physical mechanism capable of reducing the anisotropy on time scales of the order of the time it takes for a plasma parcel to travel from the reconnection region to the spacecraft, which is the maximum existence time of the anisotropic ion VDFs. For this to be the ion kinetic instabilities, they must first grow to large amplitudes and then have sufficient time to scatter the ions. Assuming that the reconnection outflow is Alfv\'enic, we estimate the travel time of the jet as $\tau_t=\delta x_t / V_{A0}$, where $V_{A0}=B_0/\sqrt{\mu_0 n_i m_i}$ is the Alfv\'en speed and $\delta x$ is the jet travel distance between the location of the spacecraft and the statistical location of the reconnection X-line at $X_{GSM}=-25~R_E$ \cite{nagai_solar_2005}. We normalize $\tau_t$ to $f_{ci0}=\omega_{ci0}/2\pi$, $\tau_t f_{ci0}=(2\pi)^{-1}\delta x_t/d_i$. The regions beyond the instabilities thresholds correspond to $\tau_t f_{ci0} \sim 10$, and slightly higher $\tau_t f_{ci0}$ values in the region bounded by the thresholds [Fig.~\ref{fig:time-scales}b]. Thus, for the instability to grow to a sufficiently large amplitude, the growth rate needs to be very large, $\gamma > 10^{-1}\omega_{ci}$, which requires substantial temperature anisotropy \cite{maruca_mms_2018}. There is a small fraction of points with anisotropies compatible with such high growth rates for $T_{i\parallel}>T_{i\perp}$, but none for $T_{i\parallel}<T_{i\perp}$. Therefore it is unlikely the fluctuations can grow to sufficiently large amplitudes.

The pitch-angle scattering time $\tau_{s}$ due to ion scale, $k\rho_{i0}\sim 1$, Alfv\'enic fluctuations can be estimated as $\tau_{s}f_{ci0}\sim ( |\delta \mathbf{B}|/B_0)^{-2}$ \cite{bale_magnetic_2009,kennel_limit_1966}. We find that $\tau_{s}$ is typically much larger than the jet travel time $\tau_t$ [Fig.~\ref{fig:time-scales}c], indicating that the wave amplitudes are not sufficiently large to scatter ions over the travel time of the jet. In particular, the time required for waves to scatter ions is $\tau_tf_{ci0}\sim\tau_{s}f_{ci0}$ and during that time, the jet would travel a distance $\delta x \sim 2\pi (|\delta \mathbf{B}|/B_0)^{-2}~d_i \sim 2\pi\times 10^{3}~d_i\approx 400~R_E$ which is much larger than the average distance from reconnection region to spacecraft $\delta x_t \approx 2.8~R_E$. This suggests that another more efficient isotropization mechanism is at play. 

The isotropization can be provided by scattering in the CS. Observed ion VDFs are close to isotropic, and $\varepsilon_i$ is minimum at the CS center [Fig.~\ref{fig:example}] suggesting efficient scattering in highly curved magnetic fields ($r_c \ll \rho_i$ where $r_c=|\mathbf{b}.\nabla \mathbf{b}|^{-1}$ is the magnetic field curvature radius with $\mathbf{b}=\mathbf{B}/|\mathbf{B}|$). Such scattering results from the breakdown of the conservation of the first adiabatic invariant $\mu=m_i v_\perp^2/2|\mathbf{B}|$ of the ion motion, which transitions to a chaotic regime, in the sense of exponentially diverging ion phase-space trajectories~\cite{lichtenberg_regular_1983}, when the curvature radius $r_c$ approaches $\rho_i$, i.e., $\kappa=\sqrt{r_c/\rho_i} \sim 1$ \cite{wagner_particle_1979,buchner_regular_1989}. Alternatively, the ion motion in the CS can be described in terms of competition between bouncing motion across the CS and gyromotion in the CS around a finite normal magnetic field $B_n$ \cite{chen_nonlinear_1992,zelenyi_quasiadiabatic_2013}. Hence, the curvature parameter $\kappa$ is equivalent to the ratio of bouncing and gyromotion time scales $\kappa=\tau_b f_{ci}$~\cite{buchner_regular_1989,drake_ion_2009}, where $\tau_b=2\pi/\omega_b=2\pi/ \sqrt{v_0 \omega_{ci0}/L}$ with $v_0$ the characteristic particle velocity and $L$ the CS thickness. We estimate $\kappa$ as the ratio of bouncing to the gyromotion time scale, assuming $L\approx 20\rho_i$ the average CS thickness from statistical studies~\cite{runov_electric_2005}. Since we observe $T_{i\parallel} > T_{i\perp}$ at the CS edges, we assume that the ion population entering and escaping from the CS consist of two cold Alfv\'enic counter streaming beams \cite{nagai_counterstreaming_2002} so that $v_0=V_{A0}$. We also estimate $\kappa$ using the multi-spacecraft methods \cite{chanteur_spatial_1998,chanteur_spatial_1998-1}. We find excellent agreement between the two estimates indicating that the assumptions of $v_0=V_{A0}$ and $L\approx 20\rho_i$ are reasonable. 

We use the measured $\kappa$ as a proxy for the regime of ion motion. The ion motion is adiabatic for $\kappa\gg 1$ with the magnetic moment $\mu$ conserved when the gyromotion is faster than the bouncing motion. For $\kappa\sim 1$, the ion motion becomes stochastic~\cite{wagner_particle_1979,buchner_regular_1989,lukin_regimes_2022}. For $\kappa\ll 1$, the ion motion is quasi-adiabatic with the generalized magnetic moment $I_z=(2\pi)^{-1}\int p_z dz$ as an integral of motion, where $(z, p_z)$ are the variables of the fast bouncing motion \cite{schindler_adiabatic_1965,sonnerup_adiabatic_1971,buchner_regular_1989}. We observe in Fig.~\ref{fig:example}g that $\kappa \leq 1$ near the CS center, which implies that the ions are in the chaotic and quasi-adiabatic regime in a thin $l\approx \rho_i$ \cite{artemyev_proton_2010} layer in the vicinity of the CS center consistent with the observed demagnetization of the ions in the CS center [Fig.~\ref{fig:example}e]. Consistent with previous observations~\cite{artemyev_radial_2016}, we find $\kappa \leq 1$ for $73\%$ of the measured ion VDFs, which implies that the majority of the ions in the reconnection jets are in the chaotic and quasi-adiabatic motion regime. 

We expect strong pitch-angle scattering in the CS in these regimes ($\kappa \leq 1$), which reduces the anisotropy and is related to the destruction of $I_z$ \cite{eastwood_consistency_1972,wagner_particle_1979,tsyganenko_pitch-angle_1982,birmingham_pitch_1984,buchner_regular_1989,ashour-abdalla_chaotic_1990}. Studies of the Hamiltonian of the ion motion in the CS showed that the characteristic phase-space diffusion time is $\tau_Df_{ci0}\sim\kappa^{-3}$~\cite{buchner_regular_1989,zelenyi_quasiadiabatic_2013}. As the solution of the diffusion equation is exponentially decaying in time, the steady ($99\%$) state isotropic VDF is reached after $t=2\log(10) \tau_D$. Since the effective ion pitch-angle scattering occurs in the thin ($l\approx \rho_i\approx L/20$) layer where $\kappa\leq 1$ \cite{artemyev_proton_2010}, the diffusion time is increased by a factor of 20 compared to scattering in the entire thick CS. These yield the effective diffusion time $\tau_{eff}\approx 40\log(10)\tau_D\sim 100 \tau_D$. From Fig.~\ref{fig:time-scales}d, we find that $\tau_D \sim 0.01 \tau_t$, so that the effective diffusion time is similar to the travel time of the observed jets, $\tau_t\approx \tau_{eff}$. In particular, we find that the effective characteristic scattering time is $\tau_{eff}\approx 18~\mathrm{s}\approx 3.3f_{ci0}^{-1}$. Thus the travel distance from the X-line necessary to scatter the ions is $\delta x \sim 2\pi\times 3.3~d_i \approx 20~d_i\approx 1.5~R_E < \delta x_t\approx 2.8~R_E$. This indicates that the interaction with the CS can efficiently scatter ions during the observed travel time of the jets. 

In summary, we have presented a statistical investigation of the ion VDFs in reconnection jets. Our results show that overall the observed ion VDFs are isotropic ($T_{i\perp} / T_{i\parallel}=0.9_{-0.3}^{+0.25}$) but are driven out of thermal equilibrium by magnetic reconnection compared to the quiet CPS. 

Our primary finding is that the phase-space diffusion due to chaotic and quasi-adiabatic ion motion in the CS is sufficiently fast to be the primary process leading to isotropization. In particular, the travel distance needed to scatter ions is only $\delta x \approx 20~d_i$. Similar isotropization scales $\sim 10~d_i$ are found in simulations \cite{aunai_proton_2011,divin_three-scale_2016}. Studies of the Hamiltonian of the ion motion~\cite{artemyev_rapid_2014,artemyev_superfast_2020} have shown that in the presence of a guide-field such as at astrophysical plasma jets, the phase-space diffusion is even faster, suggesting that for these environments the ion isotropization may be very efficient.

The unstable VDFs excite micro-instabilities resulting in enhanced wave activity, which is however too weak to explain the observed isotropization. We observe an enhancement of wave activity for plasma conditions corresponding to $\gamma \leq 10^{-2}\omega_{ci0}$ meaning that the instabilities generated waves can grow to large amplitudes after $\sim 10-100~f_{ci0}^{-1}$. Then, wave-particle interaction can isotropize ion VDFs after $\sim 10^3 f_{ci0}^{-1}$. This implies that during this time, the jets would need to travel a distance $\delta x \sim 2\pi\times 10^{3}~d_i$, much larger than the observed distance to the X-line. Thus, wave-particle scattering is too slow to explain the predominantly isotropic ion VDFs within the observed jets. We note that the wave-particle scattering is more efficient in denser environments, allowing the wave-particle interaction to scatter ions over shorter distances compared with the system size (e.g., for coronal loops $n_i\sim 1\times10^{15}~\mathrm{m}^{-3}$, $d_i\sim 7~\mathrm{m}$ \cite{reale_coronal_2014}, so that $\delta x\sim 2\pi\times 10^{3}~d_i\sim 45~\mathrm{km}\ll L_{loop}\sim 10^5~\mathrm{km}$). 

MMS data are available at the MMS Science Data Center; see Ref. \footnote{See \url{https://lasp.colorado.edu/mms/sdc/public}.}. Data analysis was performed using the \verb+pyrfu+ analysis package \footnote{See \url{https://pypi.org/project/pyrfu/}.}. 

We thank the MMS team and instrument PIs for data access and support. This work is supported by the Swedish National Space Agency Grant 139/18.

\clearpage
\begin{widetext}
\begin{center}
\textbf{\large Supplemental Material for ``Fast Ion Isotropization by Current Sheet Scattering in Magnetic Reconnection Jets''}
\end{center}
\end{widetext}
\setcounter{equation}{0}
\setcounter{figure}{0}
\setcounter{table}{0}
\setcounter{page}{1}
\makeatletter
\renewcommand{\theequation}{S\arabic{equation}}
\renewcommand{\thefigure}{S\arabic{figure}}

\section{Introduction}
We present here additional material to support the analysis presented in the Letter.

\section{Data distribution in reconnection jets and quiet plasma sheet}
\label{ssec:data_distribution}
\begin{figure}[!b]
    \centering
    \includegraphics[width=\linewidth]{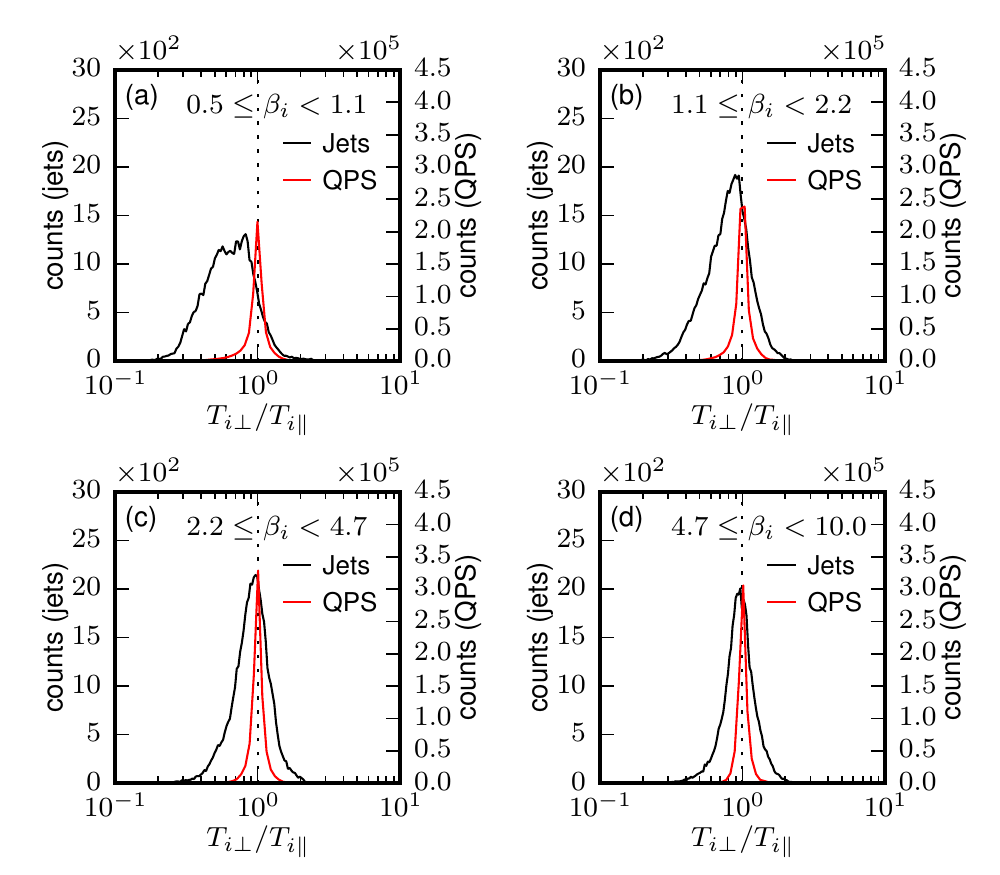}
    \caption{Histograms of ion temperature anisotropy in the reconnection jets and quiet CPS for different ranges of $\beta_i$. The black lines correspond to the reconnection jets ($\beta_i>0.5,~V_i\geq 300~\mathrm{km}~\mathrm{s}^{-1}$ as defined in~\cite{angelopoulos_statistical_1994,richard_are_2022}) and the red lines to the quiet CPS $(\beta_i>0.5,~V_i\leq 100~\mathrm{km}~\mathrm{s}^{-1})$.}
    \label{fig:figure1}
\end{figure}

As discussed in the Letter, we observe that: (a) the reconnection jets and the quiet CPS are overall isotropic, and (b) a large fraction of the VDFs in the jets show anisotropy. To better illustrate the two aspects and to better highlight the difference in the distributions for the reconnection jets ($2.6\times 10^5$ ion VDFs) versus the quiet CPS ($2.6\times 10^6$ ion VDFs), we provide in Figure~\ref{fig:figure1} histograms of $T_{i\perp}/T_{i\parallel}$ for different ranges of $\beta_i=n_ik_BT_i/P_{mag}$ where $n_i$ is the ion number density, $T_i$ is the ion temperature, and $P_{mag}=|\mathbf{B}|^2/2\mu_0$ is the magnetic pressure. The histograms of temperature anisotropies are centered around $T_{i\perp}=T_{i\parallel}$, both in the reconnection jets and the quiet CPS. However, as seen in Figure 1 in the Letter and stated in the text, the spread in the distribution of $T_{i\perp}/T_{i\parallel}$ is much larger for reconnection jets than in the quiet CPS, meaning that a large fraction of the VDFs in the jets show anisotropy.

\section{Ion non-Maxwellianity}
\subsection{Effect of time-averaging}
\label{ssec:time_avg}
To understand the relative changes of non-Maxwellianity $\varepsilon_i$ across the reconnection jet, we must ensure that our estimate of $\varepsilon_i$ is reliable in the sense that measurement uncertainties do not cause the observed relative changes. The relatively low counting statistics in the Earth's magnetotail due to the low number density, $n_i\sim0.1~\mathrm{cm}^{-3}$, results in uncertainties in the ion VDFs, $\delta f_i/f_i = 1/\sqrt{n}$, where $n$ the number of counts measured by the instrument assuming Poisson counting statistics. This can result in an artificially large non-Maxwellianity $\varepsilon$. To verify the reliability of the $\varepsilon$ estimate, we apply a test based on averaging over different windows, which have been previously applied to electron VDFs in Ref.~\cite{graham_nonmaxwellianity_2021}.

\begin{figure}[!b]
    \centering
    \includegraphics[width=\linewidth]{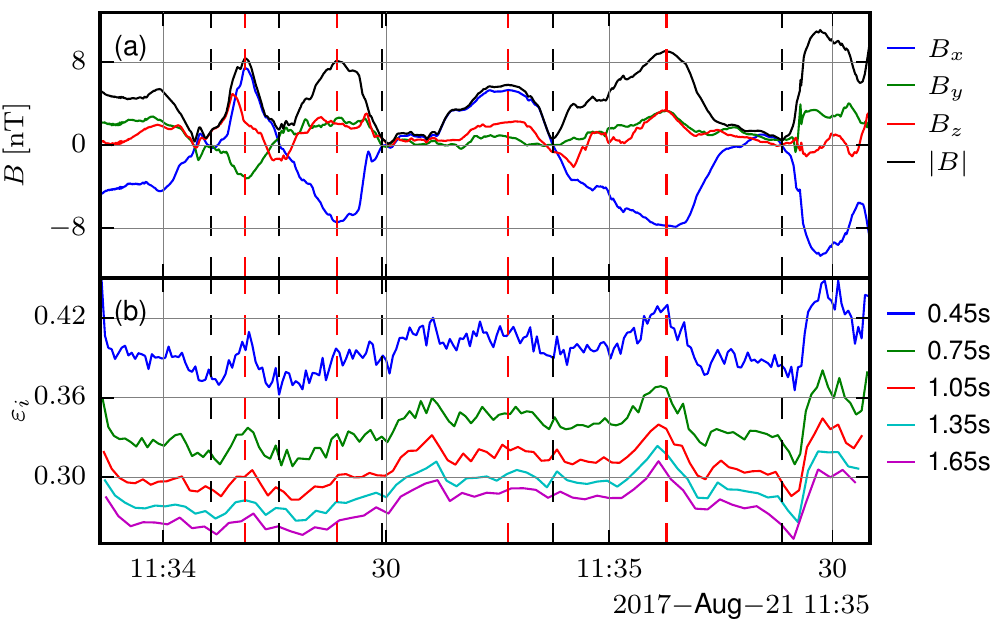}
    \caption{Comparison of the different averaging windows for the flapping event presented in Figure 2 in the Letter. (a) The magnetic field in Geocentric Solar Magnetospheric (GSM) coordinates. (b) Non-Maxwellianity $\varepsilon_i$ ion VDFs computed using different averaging windows.}
    \label{fig:figure2}
\end{figure}

To improve the counting statistics, we employ time-averaging over sliding windows of $0.45~\mathrm{s}$, $0.75~\mathrm{s}$, $1.05~\mathrm{s}$, $1.35~\mathrm{ms}$, and $1.65~\mathrm{s}$. These correspond to averaging over 3, 5, 7, 9, and 11 ion VDFs, respectively. We apply the averaging to the phase-space density $f_i(K_i,\phi,\theta)$ for each energy $K_i$, azimuthal angle $\phi$, and elevation angle $\theta$ bins of the FPI-DIS instrument~\cite{pollock_fast_2016}. Figure~\ref{fig:figure2} shows $\varepsilon_i$ for the different averaging windows for the flapping event presented in the Letter. We observe that $\varepsilon_i$ decreases with the increasing length of the averaging window. This behavior is expected as the VDFs receive more smoothing, which filters out the small velocity-space features. For the window longer than $1~\mathrm{s}$ corresponding to averaging over seven measured ion VDFs, the non-Maxwellianity converges toward similar values.

\begin{figure}[!t]
    \centering
    \includegraphics[width=\linewidth]{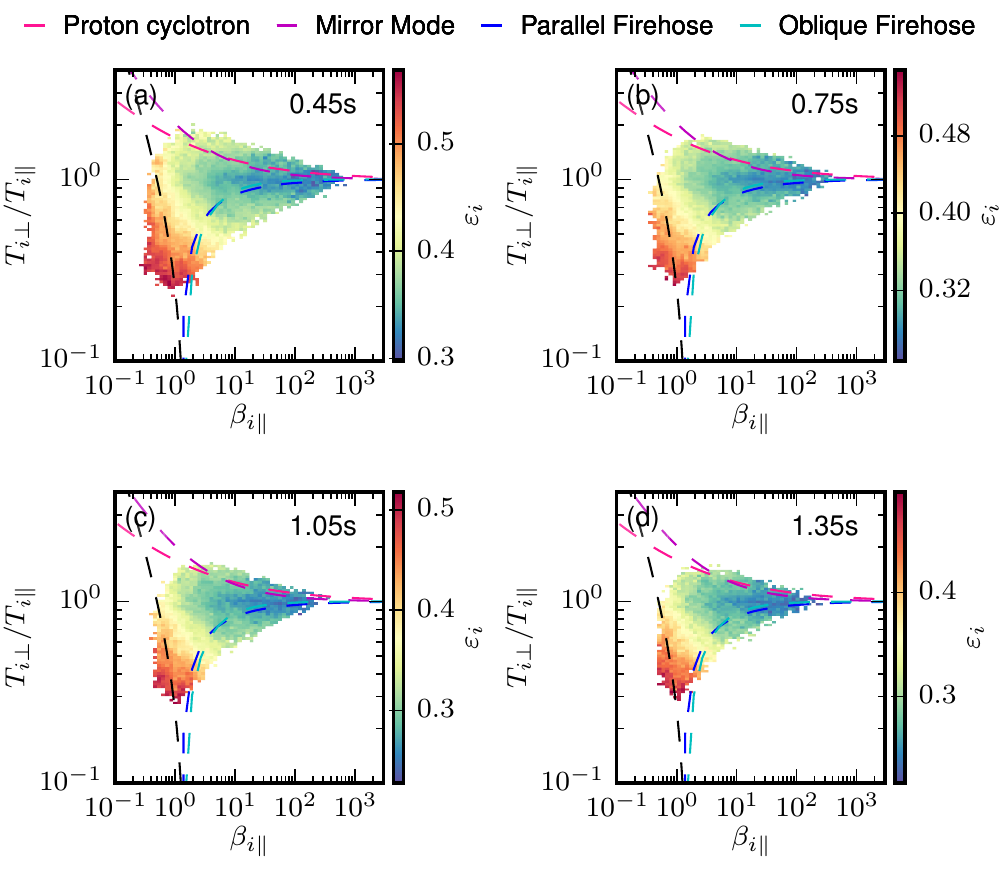}
    \caption{Comparison of the conditional average of the non-Maxwellianity $\varepsilon_i$ for different averaging windows. The black dashed line indicates $\beta_i=0.5$.}
    \label{fig:figure3}
\end{figure}

Comparing $\varepsilon_i$ estimated using the $0.45~\mathrm{s}$ and $1.65~\mathrm{s}$ window (i.e., averaging over 3 vs 11 ion VDFs), we see that the relative changes in $\varepsilon_i$ persist for different averaging, indicating that these changes are physical. In particular, the relative changes across the reconnection jet are similar for the different averaging windows. Furthermore, in Figure~\ref{fig:figure3}, we present the conditional average of $\varepsilon_i$ for different averaging windows. We see that the relative changes of $\varepsilon_i$ between the different regions of the $(\beta_{i}\parallel, T_{i\perp}/T_{i\parallel})$ phase-space persist for all averaging window sizes. This indicates that the observed changes in $\varepsilon_i$ across the reconnection jet, as described in the Letter, are physical and not caused by measurement uncertainties.

\subsection{Comparison to the quiet CPS}
To provide a reference value for the non-Maxwellianity, we estimate $\varepsilon_i$ in the quiet CPS using all ion VDFs available from 2017 to 2021 in burst mode ($150~\mathrm{ms}$). After time-averaging (see previous section), it yields $5.7\times10^4$ ion VDFs in the quiet CPS. Here we note that we use burst mode ($150~\mathrm{ms}$ time resolution) to compare the values of non-Maxwellianity using identical procedures with similar time-averaging windows. 

\begin{figure}
    \centering
    \includegraphics[width=\linewidth]{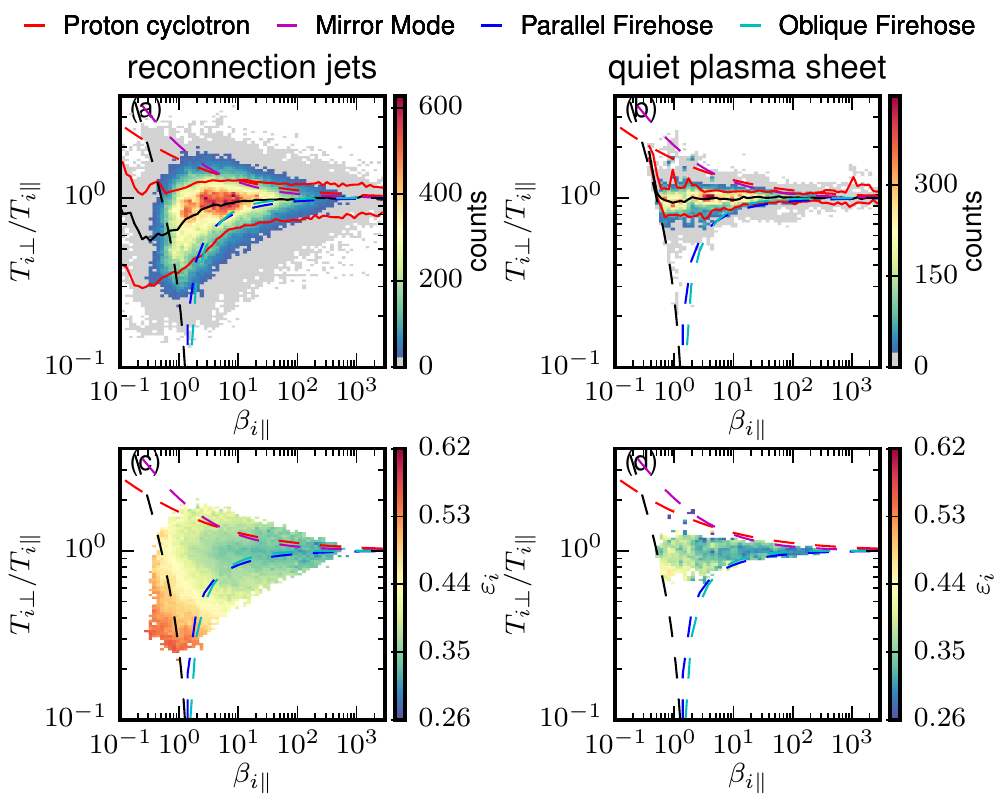}
    \caption{Comparison of data distribution and non-Maxwellianity in reconnection jets (left) and quiet CPS (right).}
    \label{fig:figure4}
\end{figure}

We present the distribution of measured ion VDFs and the non-Maxwellianity in the reconnection jets (left column) and the quiet plasma sheet (right column) in Figure~\ref{fig:figure4}. As described earlier, the temperature anisotropy in the quiet CPS is much more narrowly distributed compared to the reconnection jets. Furthermore, we find that $\varepsilon_i$ has a weak dependence on $\beta_{i\parallel}$, meaning that the quiet CPS $\varepsilon_i$ can be summarized to its mean $\langle\epsilon_i\rangle=0.35$ and its standard deviation $\sigma(\epsilon_i)=0.09$. We use $\langle\epsilon_i\rangle=0.35$ as a reference value of background non-Maxwellianity.

\bibliographystyle{apsrev4-2}
\bibliography{arxiv}

\begin{thebibliography}{64}%
\makeatletter
\providecommand \@ifxundefined [1]{%
 \@ifx{#1\undefined}
}%
\providecommand \@ifnum [1]{%
 \ifnum #1\expandafter \@firstoftwo
 \else \expandafter \@secondoftwo
 \fi
}%
\providecommand \@ifx [1]{%
 \ifx #1\expandafter \@firstoftwo
 \else \expandafter \@secondoftwo
 \fi
}%
\providecommand \natexlab [1]{#1}%
\providecommand \enquote  [1]{``#1''}%
\providecommand \bibnamefont  [1]{#1}%
\providecommand \bibfnamefont [1]{#1}%
\providecommand \citenamefont [1]{#1}%
\providecommand \href@noop [0]{\@secondoftwo}%
\providecommand \href [0]{\begingroup \@sanitize@url \@href}%
\providecommand \@href[1]{\@@startlink{#1}\@@href}%
\providecommand \@@href[1]{\endgroup#1\@@endlink}%
\providecommand \@sanitize@url [0]{\catcode `\\12\catcode `\$12\catcode
  `\&12\catcode `\#12\catcode `\^12\catcode `\_12\catcode `\%12\relax}%
\providecommand \@@startlink[1]{}%
\providecommand \@@endlink[0]{}%
\providecommand \url  [0]{\begingroup\@sanitize@url \@url }%
\providecommand \@url [1]{\endgroup\@href {#1}{\urlprefix }}%
\providecommand \urlprefix  [0]{URL }%
\providecommand \Eprint [0]{\href }%
\providecommand \doibase [0]{https://doi.org/}%
\providecommand \selectlanguage [0]{\@gobble}%
\providecommand \bibinfo  [0]{\@secondoftwo}%
\providecommand \bibfield  [0]{\@secondoftwo}%
\providecommand \translation [1]{[#1]}%
\providecommand \BibitemOpen [0]{}%
\providecommand \bibitemStop [0]{}%
\providecommand \bibitemNoStop [0]{.\EOS\space}%
\providecommand \EOS [0]{\spacefactor3000\relax}%
\providecommand \BibitemShut  [1]{\csname bibitem#1\endcsname}%
\let\auto@bib@innerbib\@empty
\bibitem [{\citenamefont {Masuda}\ \emph {et~al.}(1994)\citenamefont {Masuda},
  \citenamefont {Kosugi}, \citenamefont {Hara}, \citenamefont {Tsuneta},\ and\
  \citenamefont {Ogawara}}]{masuda_loop-top_1994}%
  \BibitemOpen
  \bibfield  {author} {\bibinfo {author} {\bibfnamefont {S.}~\bibnamefont
  {Masuda}}, \bibinfo {author} {\bibfnamefont {T.}~\bibnamefont {Kosugi}},
  \bibinfo {author} {\bibfnamefont {H.}~\bibnamefont {Hara}}, \bibinfo {author}
  {\bibfnamefont {S.}~\bibnamefont {Tsuneta}},\ and\ \bibinfo {author}
  {\bibfnamefont {Y.}~\bibnamefont {Ogawara}},\ }\href
  {https://doi.org/10.1038/371495a0} {\bibfield  {journal} {\bibinfo  {journal}
  {Nature}\ }\textbf {\bibinfo {volume} {371}},\ \bibinfo {pages} {495}
  (\bibinfo {year} {1994})}\BibitemShut {NoStop}%
\bibitem [{\citenamefont {Phan}\ \emph {et~al.}(2000)\citenamefont {Phan},
  \citenamefont {Kistler}, \citenamefont {Klecker}, \citenamefont {Haerendel},
  \citenamefont {Paschmann}, \citenamefont {Sonnerup}, \citenamefont
  {Baumjohann}, \citenamefont {Bavassano-Cattaneo}, \citenamefont {Carlson},
  \citenamefont {DiLellis}, \citenamefont {Fornacon}, \citenamefont {Frank},
  \citenamefont {Fujimoto}, \citenamefont {Georgescu}, \citenamefont {Kokubun},
  \citenamefont {Moebius}, \citenamefont {Mukai}, \citenamefont {Øieroset},
  \citenamefont {Paterson},\ and\ \citenamefont {Reme}}]{phan_extended_2000}%
  \BibitemOpen
  \bibfield  {author} {\bibinfo {author} {\bibfnamefont {T.~D.}\ \bibnamefont
  {Phan}}, \bibinfo {author} {\bibfnamefont {L.~M.}\ \bibnamefont {Kistler}},
  \bibinfo {author} {\bibfnamefont {B.}~\bibnamefont {Klecker}}, \bibinfo
  {author} {\bibfnamefont {G.}~\bibnamefont {Haerendel}}, \bibinfo {author}
  {\bibfnamefont {G.}~\bibnamefont {Paschmann}}, \bibinfo {author}
  {\bibfnamefont {B.~U.~O.}\ \bibnamefont {Sonnerup}}, \bibinfo {author}
  {\bibfnamefont {W.}~\bibnamefont {Baumjohann}}, \bibinfo {author}
  {\bibfnamefont {M.~B.}\ \bibnamefont {Bavassano-Cattaneo}}, \bibinfo {author}
  {\bibfnamefont {C.~W.}\ \bibnamefont {Carlson}}, \bibinfo {author}
  {\bibfnamefont {A.~M.}\ \bibnamefont {DiLellis}}, \bibinfo {author}
  {\bibfnamefont {K.-H.}\ \bibnamefont {Fornacon}}, \bibinfo {author}
  {\bibfnamefont {L.~A.}\ \bibnamefont {Frank}}, \bibinfo {author}
  {\bibfnamefont {M.}~\bibnamefont {Fujimoto}}, \bibinfo {author}
  {\bibfnamefont {E.}~\bibnamefont {Georgescu}}, \bibinfo {author}
  {\bibfnamefont {S.}~\bibnamefont {Kokubun}}, \bibinfo {author} {\bibfnamefont
  {E.}~\bibnamefont {Moebius}}, \bibinfo {author} {\bibfnamefont
  {T.}~\bibnamefont {Mukai}}, \bibinfo {author} {\bibfnamefont
  {M.}~\bibnamefont {Øieroset}}, \bibinfo {author} {\bibfnamefont {W.~R.}\
  \bibnamefont {Paterson}},\ and\ \bibinfo {author} {\bibfnamefont
  {H.}~\bibnamefont {Reme}},\ }\href {https://doi.org/10.1038/35009050}
  {\bibfield  {journal} {\bibinfo  {journal} {Nature}\ }\textbf {\bibinfo
  {volume} {404}},\ \bibinfo {pages} {848} (\bibinfo {year}
  {2000})}\BibitemShut {NoStop}%
\bibitem [{\citenamefont {Pudritz}\ \emph {et~al.}(2012)\citenamefont
  {Pudritz}, \citenamefont {Hardcastle},\ and\ \citenamefont
  {Gabuzda}}]{pudritz_magnetic_2012}%
  \BibitemOpen
  \bibfield  {author} {\bibinfo {author} {\bibfnamefont {R.~E.}\ \bibnamefont
  {Pudritz}}, \bibinfo {author} {\bibfnamefont {M.~J.}\ \bibnamefont
  {Hardcastle}},\ and\ \bibinfo {author} {\bibfnamefont {D.~C.}\ \bibnamefont
  {Gabuzda}},\ }\href {https://doi.org/10.1007/s11214-012-9895-z} {\bibfield
  {journal} {\bibinfo  {journal} {Space Science Reviews}\ }\textbf {\bibinfo
  {volume} {169}},\ \bibinfo {pages} {27} (\bibinfo {year} {2012})}\BibitemShut
  {NoStop}%
\bibitem [{\citenamefont {Parfrey}\ \emph {et~al.}(2019)\citenamefont
  {Parfrey}, \citenamefont {Philippov},\ and\ \citenamefont
  {Cerutti}}]{parfrey_first-principles_2019}%
  \BibitemOpen
  \bibfield  {author} {\bibinfo {author} {\bibfnamefont {K.}~\bibnamefont
  {Parfrey}}, \bibinfo {author} {\bibfnamefont {A.}~\bibnamefont {Philippov}},\
  and\ \bibinfo {author} {\bibfnamefont {B.}~\bibnamefont {Cerutti}},\ }\href
  {https://doi.org/10.1103/PhysRevLett.122.035101} {\bibfield  {journal}
  {\bibinfo  {journal} {Physical Review Letters}\ }\textbf {\bibinfo {volume}
  {122}},\ \bibinfo {pages} {035101} (\bibinfo {year} {2019})}\BibitemShut
  {NoStop}%
\bibitem [{\citenamefont {Biskamp}(2000)}]{biskamp_magnetic_2000}%
  \BibitemOpen
  \bibfield  {author} {\bibinfo {author} {\bibfnamefont {D.}~\bibnamefont
  {Biskamp}},\ }\href {https://doi.org/10.1017/CBO9780511599958} {\emph
  {\bibinfo {title} {Magnetic reconnection in plasmas}}},\ Cambridge
  {Monographs} on {Plasma} {Physics}\ (\bibinfo  {publisher} {Cambridge
  University Press},\ \bibinfo {address} {Cambridge},\ \bibinfo {year}
  {2000})\BibitemShut {NoStop}%
\bibitem [{\citenamefont {Yamada}\ \emph {et~al.}(2010)\citenamefont {Yamada},
  \citenamefont {Kulsrud},\ and\ \citenamefont {Ji}}]{yamada_magnetic_2010}%
  \BibitemOpen
  \bibfield  {author} {\bibinfo {author} {\bibfnamefont {M.}~\bibnamefont
  {Yamada}}, \bibinfo {author} {\bibfnamefont {R.}~\bibnamefont {Kulsrud}},\
  and\ \bibinfo {author} {\bibfnamefont {H.}~\bibnamefont {Ji}},\ }\href
  {https://doi.org/10.1103/RevModPhys.82.603} {\bibfield  {journal} {\bibinfo
  {journal} {Reviews of Modern Physics}\ }\textbf {\bibinfo {volume} {82}},\
  \bibinfo {pages} {603} (\bibinfo {year} {2010})}\BibitemShut {NoStop}%
\bibitem [{\citenamefont {Hoshino}\ \emph {et~al.}(1998)\citenamefont
  {Hoshino}, \citenamefont {Mukai}, \citenamefont {Yamamoto},\ and\
  \citenamefont {Kokubun}}]{hoshino_ion_1998}%
  \BibitemOpen
  \bibfield  {author} {\bibinfo {author} {\bibfnamefont {M.}~\bibnamefont
  {Hoshino}}, \bibinfo {author} {\bibfnamefont {T.}~\bibnamefont {Mukai}},
  \bibinfo {author} {\bibfnamefont {T.}~\bibnamefont {Yamamoto}},\ and\
  \bibinfo {author} {\bibfnamefont {S.}~\bibnamefont {Kokubun}},\ }\href
  {https://doi.org/10.1029/97JA01785} {\bibfield  {journal} {\bibinfo
  {journal} {Journal of Geophysical Research: Space Physics}\ }\textbf
  {\bibinfo {volume} {103}},\ \bibinfo {pages} {4509} (\bibinfo {year}
  {1998})}\BibitemShut {NoStop}%
\bibitem [{\citenamefont {Nagai}\ \emph {et~al.}(2002)\citenamefont {Nagai},
  \citenamefont {Nakamura}, \citenamefont {Shinohara}, \citenamefont
  {Fujimoto}, \citenamefont {Saito},\ and\ \citenamefont
  {Mukai}}]{nagai_counterstreaming_2002}%
  \BibitemOpen
  \bibfield  {author} {\bibinfo {author} {\bibfnamefont {T.}~\bibnamefont
  {Nagai}}, \bibinfo {author} {\bibfnamefont {M.}~\bibnamefont {Nakamura}},
  \bibinfo {author} {\bibfnamefont {I.}~\bibnamefont {Shinohara}}, \bibinfo
  {author} {\bibfnamefont {M.}~\bibnamefont {Fujimoto}}, \bibinfo {author}
  {\bibfnamefont {Y.}~\bibnamefont {Saito}},\ and\ \bibinfo {author}
  {\bibfnamefont {T.}~\bibnamefont {Mukai}},\ }\href
  {https://doi.org/10.1063/1.1499117} {\bibfield  {journal} {\bibinfo
  {journal} {Physics of Plasmas}\ }\textbf {\bibinfo {volume} {9}},\ \bibinfo
  {pages} {3705} (\bibinfo {year} {2002})}\BibitemShut {NoStop}%
\bibitem [{\citenamefont {Eastwood}\ \emph {et~al.}(2015)\citenamefont
  {Eastwood}, \citenamefont {Goldman}, \citenamefont {Hietala}, \citenamefont
  {Newman}, \citenamefont {Mistry},\ and\ \citenamefont
  {Lapenta}}]{eastwood_ion_2015}%
  \BibitemOpen
  \bibfield  {author} {\bibinfo {author} {\bibfnamefont {J.~P.}\ \bibnamefont
  {Eastwood}}, \bibinfo {author} {\bibfnamefont {M.~V.}\ \bibnamefont
  {Goldman}}, \bibinfo {author} {\bibfnamefont {H.}~\bibnamefont {Hietala}},
  \bibinfo {author} {\bibfnamefont {D.~L.}\ \bibnamefont {Newman}}, \bibinfo
  {author} {\bibfnamefont {R.}~\bibnamefont {Mistry}},\ and\ \bibinfo {author}
  {\bibfnamefont {G.}~\bibnamefont {Lapenta}},\ }\href
  {https://doi.org/10.1002/2014JA020516} {\bibfield  {journal} {\bibinfo
  {journal} {Journal of Geophysical Research: Space Physics}\ }\textbf
  {\bibinfo {volume} {120}},\ \bibinfo {pages} {511} (\bibinfo {year}
  {2015})}\BibitemShut {NoStop}%
\bibitem [{\citenamefont {Liu}\ \emph {et~al.}(2019)\citenamefont {Liu},
  \citenamefont {Vaivads}, \citenamefont {Graham}, \citenamefont
  {Khotyaintsev}, \citenamefont {Fu}, \citenamefont {Johlander}, \citenamefont
  {André},\ and\ \citenamefont {Giles}}]{liu_ionbeamdriven_2019}%
  \BibitemOpen
  \bibfield  {author} {\bibinfo {author} {\bibfnamefont {C.~M.}\ \bibnamefont
  {Liu}}, \bibinfo {author} {\bibfnamefont {A.}~\bibnamefont {Vaivads}},
  \bibinfo {author} {\bibfnamefont {D.~B.}\ \bibnamefont {Graham}}, \bibinfo
  {author} {\bibfnamefont {Y.~V.}\ \bibnamefont {Khotyaintsev}}, \bibinfo
  {author} {\bibfnamefont {H.~S.}\ \bibnamefont {Fu}}, \bibinfo {author}
  {\bibfnamefont {A.}~\bibnamefont {Johlander}}, \bibinfo {author}
  {\bibfnamefont {M.}~\bibnamefont {André}},\ and\ \bibinfo {author}
  {\bibfnamefont {B.~L.}\ \bibnamefont {Giles}},\ }\href
  {https://doi.org/10.1029/2019GL085419} {\bibfield  {journal} {\bibinfo
  {journal} {Geophysical Research Letters}\ }\textbf {\bibinfo {volume} {46}},\
  \bibinfo {pages} {12702} (\bibinfo {year} {2019})}\BibitemShut {NoStop}%
\bibitem [{\citenamefont {Drake}\ \emph {et~al.}(2009)\citenamefont {Drake},
  \citenamefont {Swisdak}, \citenamefont {Phan}, \citenamefont {Cassak},
  \citenamefont {Shay}, \citenamefont {Lepri}, \citenamefont {Lin},
  \citenamefont {Quataert},\ and\ \citenamefont {Zurbuchen}}]{drake_ion_2009}%
  \BibitemOpen
  \bibfield  {author} {\bibinfo {author} {\bibfnamefont {J.~F.}\ \bibnamefont
  {Drake}}, \bibinfo {author} {\bibfnamefont {M.}~\bibnamefont {Swisdak}},
  \bibinfo {author} {\bibfnamefont {T.~D.}\ \bibnamefont {Phan}}, \bibinfo
  {author} {\bibfnamefont {P.~A.}\ \bibnamefont {Cassak}}, \bibinfo {author}
  {\bibfnamefont {M.~A.}\ \bibnamefont {Shay}}, \bibinfo {author}
  {\bibfnamefont {S.~T.}\ \bibnamefont {Lepri}}, \bibinfo {author}
  {\bibfnamefont {R.~P.}\ \bibnamefont {Lin}}, \bibinfo {author} {\bibfnamefont
  {E.}~\bibnamefont {Quataert}},\ and\ \bibinfo {author} {\bibfnamefont
  {T.~H.}\ \bibnamefont {Zurbuchen}},\ }\href
  {https://doi.org/10.1029/2008JA013701} {\bibfield  {journal} {\bibinfo
  {journal} {Journal of Geophysical Research: Space Physics}\ }\textbf
  {\bibinfo {volume} {114}},\ \bibinfo {pages} {A05111} (\bibinfo {year}
  {2009})}\BibitemShut {NoStop}%
\bibitem [{\citenamefont {Hietala}\ \emph {et~al.}(2015)\citenamefont
  {Hietala}, \citenamefont {Drake}, \citenamefont {Phan}, \citenamefont
  {Eastwood},\ and\ \citenamefont {McFadden}}]{hietala_ion_2015}%
  \BibitemOpen
  \bibfield  {author} {\bibinfo {author} {\bibfnamefont {H.}~\bibnamefont
  {Hietala}}, \bibinfo {author} {\bibfnamefont {J.~F.}\ \bibnamefont {Drake}},
  \bibinfo {author} {\bibfnamefont {T.~D.}\ \bibnamefont {Phan}}, \bibinfo
  {author} {\bibfnamefont {J.~P.}\ \bibnamefont {Eastwood}},\ and\ \bibinfo
  {author} {\bibfnamefont {J.~P.}\ \bibnamefont {McFadden}},\ }\href
  {https://doi.org/10.1002/2015GL065168} {\bibfield  {journal} {\bibinfo
  {journal} {Geophysical Research Letters}\ }\textbf {\bibinfo {volume} {42}},\
  \bibinfo {pages} {7239} (\bibinfo {year} {2015})}\BibitemShut {NoStop}%
\bibitem [{\citenamefont {Gary}(1993)}]{gary_theory_1993}%
  \BibitemOpen
  \bibfield  {author} {\bibinfo {author} {\bibfnamefont {S.~P.}\ \bibnamefont
  {Gary}},\ }\href@noop {} {\emph {\bibinfo {title} {Theory of space plasma
  microinstabilities}}},\ Cambridge atmospheric and space science series\
  (\bibinfo  {publisher} {Cambridge University Press},\ \bibinfo {address}
  {Cambridge},\ \bibinfo {year} {1993})\BibitemShut {NoStop}%
\bibitem [{\citenamefont {Vörös}(2011)}]{voros_magnetic_2011}%
  \BibitemOpen
  \bibfield  {author} {\bibinfo {author} {\bibfnamefont {Z.}~\bibnamefont
  {Vörös}},\ }\href {https://doi.org/10.5194/npg-18-861-2011} {\bibfield
  {journal} {\bibinfo  {journal} {Nonlinear Processes in Geophysics}\ }\textbf
  {\bibinfo {volume} {18}},\ \bibinfo {pages} {861} (\bibinfo {year}
  {2011})}\BibitemShut {NoStop}%
\bibitem [{\citenamefont {Wu}\ \emph {et~al.}(2013)\citenamefont {Wu},
  \citenamefont {Volwerk}, \citenamefont {Lu}, \citenamefont {Vörös},
  \citenamefont {Nakamura},\ and\ \citenamefont {Zhang}}]{wu_proton_2013}%
  \BibitemOpen
  \bibfield  {author} {\bibinfo {author} {\bibfnamefont {M.}~\bibnamefont
  {Wu}}, \bibinfo {author} {\bibfnamefont {M.}~\bibnamefont {Volwerk}},
  \bibinfo {author} {\bibfnamefont {Q.}~\bibnamefont {Lu}}, \bibinfo {author}
  {\bibfnamefont {Z.}~\bibnamefont {Vörös}}, \bibinfo {author} {\bibfnamefont
  {R.}~\bibnamefont {Nakamura}},\ and\ \bibinfo {author} {\bibfnamefont
  {T.}~\bibnamefont {Zhang}},\ }\href {https://doi.org/10.1002/jgra.50451}
  {\bibfield  {journal} {\bibinfo  {journal} {Journal of Geophysical Research:
  Space Physics}\ }\textbf {\bibinfo {volume} {118}},\ \bibinfo {pages} {4875}
  (\bibinfo {year} {2013})}\BibitemShut {NoStop}%
\bibitem [{\citenamefont {Kunz}\ \emph {et~al.}(2014)\citenamefont {Kunz},
  \citenamefont {Schekochihin},\ and\ \citenamefont
  {Stone}}]{kunz_firehose_2014}%
  \BibitemOpen
  \bibfield  {author} {\bibinfo {author} {\bibfnamefont {M.~W.}\ \bibnamefont
  {Kunz}}, \bibinfo {author} {\bibfnamefont {A.~A.}\ \bibnamefont
  {Schekochihin}},\ and\ \bibinfo {author} {\bibfnamefont {J.~M.}\ \bibnamefont
  {Stone}},\ }\href {https://doi.org/10.1103/PhysRevLett.112.205003} {\bibfield
   {journal} {\bibinfo  {journal} {Physical Review Letters}\ }\textbf {\bibinfo
  {volume} {112}},\ \bibinfo {pages} {205003} (\bibinfo {year}
  {2014})}\BibitemShut {NoStop}%
\bibitem [{\citenamefont {Eastwood}(1972)}]{eastwood_consistency_1972}%
  \BibitemOpen
  \bibfield  {author} {\bibinfo {author} {\bibfnamefont {J.}~\bibnamefont
  {Eastwood}},\ }\href {https://doi.org/10.1016/0032-0633(72)90182-1}
  {\bibfield  {journal} {\bibinfo  {journal} {Planetary and Space Science}\
  }\textbf {\bibinfo {volume} {20}},\ \bibinfo {pages} {1555} (\bibinfo {year}
  {1972})}\BibitemShut {NoStop}%
\bibitem [{\citenamefont {Wagner}\ \emph {et~al.}(1979)\citenamefont {Wagner},
  \citenamefont {Kan},\ and\ \citenamefont {Akasofu}}]{wagner_particle_1979}%
  \BibitemOpen
  \bibfield  {author} {\bibinfo {author} {\bibfnamefont {J.~S.}\ \bibnamefont
  {Wagner}}, \bibinfo {author} {\bibfnamefont {J.~R.}\ \bibnamefont {Kan}},\
  and\ \bibinfo {author} {\bibfnamefont {S.-I.}\ \bibnamefont {Akasofu}},\
  }\href {https://doi.org/10.1029/JA084iA03p00891} {\bibfield  {journal}
  {\bibinfo  {journal} {Journal of Geophysical Research}\ }\textbf {\bibinfo
  {volume} {84}},\ \bibinfo {pages} {891} (\bibinfo {year} {1979})}\BibitemShut
  {NoStop}%
\bibitem [{\citenamefont {Tsyganenko}(1982)}]{tsyganenko_pitch-angle_1982}%
  \BibitemOpen
  \bibfield  {author} {\bibinfo {author} {\bibfnamefont {N.}~\bibnamefont
  {Tsyganenko}},\ }\href {https://doi.org/10.1016/0032-0633(82)90052-6}
  {\bibfield  {journal} {\bibinfo  {journal} {Planetary and Space Science}\
  }\textbf {\bibinfo {volume} {30}},\ \bibinfo {pages} {433} (\bibinfo {year}
  {1982})}\BibitemShut {NoStop}%
\bibitem [{\citenamefont {Birmingham}(1984)}]{birmingham_pitch_1984}%
  \BibitemOpen
  \bibfield  {author} {\bibinfo {author} {\bibfnamefont {T.~J.}\ \bibnamefont
  {Birmingham}},\ }\href {https://doi.org/10.1029/JA089iA05p02699} {\bibfield
  {journal} {\bibinfo  {journal} {Journal of Geophysical Research}\ }\textbf
  {\bibinfo {volume} {89}},\ \bibinfo {pages} {2699} (\bibinfo {year}
  {1984})}\BibitemShut {NoStop}%
\bibitem [{\citenamefont {Büchner}\ and\ \citenamefont
  {Zelenyi}(1989)}]{buchner_regular_1989}%
  \BibitemOpen
  \bibfield  {author} {\bibinfo {author} {\bibfnamefont {J.}~\bibnamefont
  {Büchner}}\ and\ \bibinfo {author} {\bibfnamefont {L.~M.}\ \bibnamefont
  {Zelenyi}},\ }\href {https://doi.org/10.1029/JA094iA09p11821} {\bibfield
  {journal} {\bibinfo  {journal} {Journal of Geophysical Research}\ }\textbf
  {\bibinfo {volume} {94}},\ \bibinfo {pages} {11821} (\bibinfo {year}
  {1989})}\BibitemShut {NoStop}%
\bibitem [{\citenamefont {Ashour-Abdalla}\ \emph {et~al.}(1990)\citenamefont
  {Ashour-Abdalla}, \citenamefont {Berchem}, \citenamefont {Büchner},\ and\
  \citenamefont {Zelenyi}}]{ashour-abdalla_chaotic_1990}%
  \BibitemOpen
  \bibfield  {author} {\bibinfo {author} {\bibfnamefont {M.}~\bibnamefont
  {Ashour-Abdalla}}, \bibinfo {author} {\bibfnamefont {J.}~\bibnamefont
  {Berchem}}, \bibinfo {author} {\bibfnamefont {J.}~\bibnamefont {Büchner}},\
  and\ \bibinfo {author} {\bibfnamefont {L.~M.}\ \bibnamefont {Zelenyi}},\
  }\href {https://doi.org/10.1029/GL017i013p02317} {\bibfield  {journal}
  {\bibinfo  {journal} {Geophysical Research Letters}\ }\textbf {\bibinfo
  {volume} {17}},\ \bibinfo {pages} {2317} (\bibinfo {year}
  {1990})}\BibitemShut {NoStop}%
\bibitem [{\citenamefont {Burch}\ \emph {et~al.}(2016)\citenamefont {Burch},
  \citenamefont {Moore}, \citenamefont {Torbert},\ and\ \citenamefont
  {Giles}}]{burch_magnetospheric_2016}%
  \BibitemOpen
  \bibfield  {author} {\bibinfo {author} {\bibfnamefont {J.~L.}\ \bibnamefont
  {Burch}}, \bibinfo {author} {\bibfnamefont {T.~E.}\ \bibnamefont {Moore}},
  \bibinfo {author} {\bibfnamefont {R.~B.}\ \bibnamefont {Torbert}},\ and\
  \bibinfo {author} {\bibfnamefont {B.~L.}\ \bibnamefont {Giles}},\ }\href
  {https://doi.org/10.1007/s11214-015-0164-9} {\bibfield  {journal} {\bibinfo
  {journal} {Space Science Reviews}\ }\textbf {\bibinfo {volume} {199}},\
  \bibinfo {pages} {5} (\bibinfo {year} {2016})}\BibitemShut {NoStop}%
\bibitem [{\citenamefont {Angelopoulos}\ \emph {et~al.}(1994)\citenamefont
  {Angelopoulos}, \citenamefont {Kennel}, \citenamefont {Coroniti},
  \citenamefont {Pellat}, \citenamefont {Kivelson}, \citenamefont {Walker},
  \citenamefont {Russell}, \citenamefont {Baumjohann}, \citenamefont
  {Feldman},\ and\ \citenamefont {Gosling}}]{angelopoulos_statistical_1994}%
  \BibitemOpen
  \bibfield  {author} {\bibinfo {author} {\bibfnamefont {V.}~\bibnamefont
  {Angelopoulos}}, \bibinfo {author} {\bibfnamefont {C.~F.}\ \bibnamefont
  {Kennel}}, \bibinfo {author} {\bibfnamefont {F.~V.}\ \bibnamefont
  {Coroniti}}, \bibinfo {author} {\bibfnamefont {R.}~\bibnamefont {Pellat}},
  \bibinfo {author} {\bibfnamefont {M.~G.}\ \bibnamefont {Kivelson}}, \bibinfo
  {author} {\bibfnamefont {R.~J.}\ \bibnamefont {Walker}}, \bibinfo {author}
  {\bibfnamefont {C.~T.}\ \bibnamefont {Russell}}, \bibinfo {author}
  {\bibfnamefont {W.}~\bibnamefont {Baumjohann}}, \bibinfo {author}
  {\bibfnamefont {W.~C.}\ \bibnamefont {Feldman}},\ and\ \bibinfo {author}
  {\bibfnamefont {J.~T.}\ \bibnamefont {Gosling}},\ }\href
  {https://doi.org/10.1029/94JA01263} {\bibfield  {journal} {\bibinfo
  {journal} {Journal of Geophysical Research}\ }\textbf {\bibinfo {volume}
  {99}},\ \bibinfo {pages} {21257} (\bibinfo {year} {1994})}\BibitemShut
  {NoStop}%
\bibitem [{\citenamefont {Richard}\ \emph {et~al.}(2022)\citenamefont
  {Richard}, \citenamefont {Khotyaintsev}, \citenamefont {Graham},\ and\
  \citenamefont {Russell}}]{richard_are_2022}%
  \BibitemOpen
  \bibfield  {author} {\bibinfo {author} {\bibfnamefont {L.}~\bibnamefont
  {Richard}}, \bibinfo {author} {\bibfnamefont {Y.~V.}\ \bibnamefont
  {Khotyaintsev}}, \bibinfo {author} {\bibfnamefont {D.~B.}\ \bibnamefont
  {Graham}},\ and\ \bibinfo {author} {\bibfnamefont {C.~T.}\ \bibnamefont
  {Russell}},\ }\href {https://doi.org/10.1029/2022GL101693} {\bibfield
  {journal} {\bibinfo  {journal} {Geophysical Research Letters}\ }\textbf
  {\bibinfo {volume} {49}},\ \bibinfo {pages} {e2022GL101693} (\bibinfo {year}
  {2022})}\BibitemShut {NoStop}%
\bibitem [{\citenamefont {Pritchett}\ and\ \citenamefont
  {Coroniti}(2010)}]{pritchett_kinetic_2010}%
  \BibitemOpen
  \bibfield  {author} {\bibinfo {author} {\bibfnamefont {P.~L.}\ \bibnamefont
  {Pritchett}}\ and\ \bibinfo {author} {\bibfnamefont {F.~V.}\ \bibnamefont
  {Coroniti}},\ }\href {https://doi.org/10.1029/2009JA014752} {\bibfield
  {journal} {\bibinfo  {journal} {Journal of Geophysical Research: Space
  Physics}\ }\textbf {\bibinfo {volume} {115}},\ \bibinfo {pages} {A06301}
  (\bibinfo {year} {2010})}\BibitemShut {NoStop}%
\bibitem [{\citenamefont {Nagai}\ \emph {et~al.}(2005)\citenamefont {Nagai},
  \citenamefont {Fujimoto}, \citenamefont {Nakamura}, \citenamefont
  {Baumjohann}, \citenamefont {Ieda}, \citenamefont {Shinohara}, \citenamefont
  {Machida}, \citenamefont {Saito},\ and\ \citenamefont
  {Mukai}}]{nagai_solar_2005}%
  \BibitemOpen
  \bibfield  {author} {\bibinfo {author} {\bibfnamefont {T.}~\bibnamefont
  {Nagai}}, \bibinfo {author} {\bibfnamefont {M.}~\bibnamefont {Fujimoto}},
  \bibinfo {author} {\bibfnamefont {R.}~\bibnamefont {Nakamura}}, \bibinfo
  {author} {\bibfnamefont {W.}~\bibnamefont {Baumjohann}}, \bibinfo {author}
  {\bibfnamefont {A.}~\bibnamefont {Ieda}}, \bibinfo {author} {\bibfnamefont
  {I.}~\bibnamefont {Shinohara}}, \bibinfo {author} {\bibfnamefont
  {S.}~\bibnamefont {Machida}}, \bibinfo {author} {\bibfnamefont
  {Y.}~\bibnamefont {Saito}},\ and\ \bibinfo {author} {\bibfnamefont
  {T.}~\bibnamefont {Mukai}},\ }\href {https://doi.org/10.1029/2005JA011207}
  {\bibfield  {journal} {\bibinfo  {journal} {Journal of Geophysical Research:
  Space Physics}\ }\textbf {\bibinfo {volume} {110}},\ \bibinfo {pages}
  {A09208} (\bibinfo {year} {2005})}\BibitemShut {NoStop}%
\bibitem [{\citenamefont {Russell}\ \emph {et~al.}(2016)\citenamefont
  {Russell}, \citenamefont {Anderson}, \citenamefont {Baumjohann},
  \citenamefont {Bromund}, \citenamefont {Dearborn}, \citenamefont {Fischer},
  \citenamefont {Le}, \citenamefont {Leinweber}, \citenamefont {Leneman},
  \citenamefont {Magnes}, \citenamefont {Means}, \citenamefont {Moldwin},
  \citenamefont {Nakamura}, \citenamefont {Pierce}, \citenamefont {Plaschke},
  \citenamefont {Rowe}, \citenamefont {Slavin}, \citenamefont {Strangeway},
  \citenamefont {Torbert}, \citenamefont {Hagen}, \citenamefont {Jernej},
  \citenamefont {Valavanoglou},\ and\ \citenamefont
  {Richter}}]{russell_magnetospheric_2016}%
  \BibitemOpen
  \bibfield  {author} {\bibinfo {author} {\bibfnamefont {C.~T.}\ \bibnamefont
  {Russell}}, \bibinfo {author} {\bibfnamefont {B.~J.}\ \bibnamefont
  {Anderson}}, \bibinfo {author} {\bibfnamefont {W.}~\bibnamefont
  {Baumjohann}}, \bibinfo {author} {\bibfnamefont {K.~R.}\ \bibnamefont
  {Bromund}}, \bibinfo {author} {\bibfnamefont {D.}~\bibnamefont {Dearborn}},
  \bibinfo {author} {\bibfnamefont {D.}~\bibnamefont {Fischer}}, \bibinfo
  {author} {\bibfnamefont {G.}~\bibnamefont {Le}}, \bibinfo {author}
  {\bibfnamefont {H.~K.}\ \bibnamefont {Leinweber}}, \bibinfo {author}
  {\bibfnamefont {D.}~\bibnamefont {Leneman}}, \bibinfo {author} {\bibfnamefont
  {W.}~\bibnamefont {Magnes}}, \bibinfo {author} {\bibfnamefont {J.~D.}\
  \bibnamefont {Means}}, \bibinfo {author} {\bibfnamefont {M.~B.}\ \bibnamefont
  {Moldwin}}, \bibinfo {author} {\bibfnamefont {R.}~\bibnamefont {Nakamura}},
  \bibinfo {author} {\bibfnamefont {D.}~\bibnamefont {Pierce}}, \bibinfo
  {author} {\bibfnamefont {F.}~\bibnamefont {Plaschke}}, \bibinfo {author}
  {\bibfnamefont {K.~M.}\ \bibnamefont {Rowe}}, \bibinfo {author}
  {\bibfnamefont {J.~A.}\ \bibnamefont {Slavin}}, \bibinfo {author}
  {\bibfnamefont {R.~J.}\ \bibnamefont {Strangeway}}, \bibinfo {author}
  {\bibfnamefont {R.}~\bibnamefont {Torbert}}, \bibinfo {author} {\bibfnamefont
  {C.}~\bibnamefont {Hagen}}, \bibinfo {author} {\bibfnamefont
  {I.}~\bibnamefont {Jernej}}, \bibinfo {author} {\bibfnamefont
  {A.}~\bibnamefont {Valavanoglou}},\ and\ \bibinfo {author} {\bibfnamefont
  {I.}~\bibnamefont {Richter}},\ }\href
  {https://doi.org/10.1007/s11214-014-0057-3} {\bibfield  {journal} {\bibinfo
  {journal} {Space Science Reviews}\ }\textbf {\bibinfo {volume} {199}},\
  \bibinfo {pages} {189} (\bibinfo {year} {2016})}\BibitemShut {NoStop}%
\bibitem [{\citenamefont {Pollock}\ \emph {et~al.}(2016)\citenamefont
  {Pollock}, \citenamefont {Moore}, \citenamefont {Jacques}, \citenamefont
  {Burch}, \citenamefont {Gliese}, \citenamefont {Saito}, \citenamefont
  {Omoto}, \citenamefont {Avanov}, \citenamefont {Barrie}, \citenamefont
  {Coffey}, \citenamefont {Dorelli}, \citenamefont {Gershman}, \citenamefont
  {Giles}, \citenamefont {Rosnack}, \citenamefont {Salo}, \citenamefont
  {Yokota}, \citenamefont {Adrian}, \citenamefont {Aoustin}, \citenamefont
  {Auletti}, \citenamefont {Aung}, \citenamefont {Bigio}, \citenamefont {Cao},
  \citenamefont {Chandler}, \citenamefont {Chornay}, \citenamefont {Christian},
  \citenamefont {Clark}, \citenamefont {Collinson}, \citenamefont {Corris},
  \citenamefont {De Los Santos}, \citenamefont {Devlin}, \citenamefont
  {Diaz}, \citenamefont {Dickerson}, \citenamefont {Dickson}, \citenamefont
  {Diekmann}, \citenamefont {Diggs}, \citenamefont {Duncan}, \citenamefont
  {Figueroa-Vinas}, \citenamefont {Firman}, \citenamefont {Freeman},
  \citenamefont {Galassi}, \citenamefont {Garcia}, \citenamefont {Goodhart},
  \citenamefont {Guererro}, \citenamefont {Hageman}, \citenamefont {Hanley},
  \citenamefont {Hemminger}, \citenamefont {Holland}, \citenamefont {Hutchins},
  \citenamefont {James}, \citenamefont {Jones}, \citenamefont {Kreisler},
  \citenamefont {Kujawski}, \citenamefont {Lavu}, \citenamefont {Lobell},
  \citenamefont {LeCompte}, \citenamefont {Lukemire}, \citenamefont
  {MacDonald}, \citenamefont {Mariano}, \citenamefont {Mukai}, \citenamefont
  {Narayanan}, \citenamefont {Nguyan}, \citenamefont {Onizuka}, \citenamefont
  {Paterson}, \citenamefont {Persyn}, \citenamefont {Piepgrass}, \citenamefont
  {Cheney}, \citenamefont {Rager}, \citenamefont {Raghuram}, \citenamefont
  {Ramil}, \citenamefont {Reichenthal}, \citenamefont {Rodriguez},
  \citenamefont {Rouzaud}, \citenamefont {Rucker}, \citenamefont {Saito},
  \citenamefont {Samara}, \citenamefont {Sauvaud}, \citenamefont {Schuster},
  \citenamefont {Shappirio}, \citenamefont {Shelton}, \citenamefont {Sher},
  \citenamefont {Smith}, \citenamefont {Smith}, \citenamefont {Smith},
  \citenamefont {Steinfeld}, \citenamefont {Szymkiewicz}, \citenamefont
  {Tanimoto}, \citenamefont {Taylor}, \citenamefont {Tucker}, \citenamefont
  {Tull}, \citenamefont {Uhl}, \citenamefont {Vloet}, \citenamefont {Walpole},
  \citenamefont {Weidner}, \citenamefont {White}, \citenamefont {Winkert},
  \citenamefont {Yeh},\ and\ \citenamefont {Zeuch}}]{pollock_fast_2016}%
  \BibitemOpen
  \bibfield  {author} {\bibinfo {author} {\bibfnamefont {C.}~\bibnamefont
  {Pollock}}, \bibinfo {author} {\bibfnamefont {T.}~\bibnamefont {Moore}},
  \bibinfo {author} {\bibfnamefont {A.}~\bibnamefont {Jacques}}, \bibinfo
  {author} {\bibfnamefont {J.}~\bibnamefont {Burch}}, \bibinfo {author}
  {\bibfnamefont {U.}~\bibnamefont {Gliese}}, \bibinfo {author} {\bibfnamefont
  {Y.}~\bibnamefont {Saito}}, \bibinfo {author} {\bibfnamefont
  {T.}~\bibnamefont {Omoto}}, \bibinfo {author} {\bibfnamefont
  {L.}~\bibnamefont {Avanov}}, \bibinfo {author} {\bibfnamefont
  {A.}~\bibnamefont {Barrie}}, \bibinfo {author} {\bibfnamefont
  {V.}~\bibnamefont {Coffey}}, \bibinfo {author} {\bibfnamefont
  {J.}~\bibnamefont {Dorelli}}, \bibinfo {author} {\bibfnamefont
  {D.}~\bibnamefont {Gershman}}, \bibinfo {author} {\bibfnamefont
  {B.}~\bibnamefont {Giles}}, \bibinfo {author} {\bibfnamefont
  {T.}~\bibnamefont {Rosnack}}, \bibinfo {author} {\bibfnamefont
  {C.}~\bibnamefont {Salo}}, \bibinfo {author} {\bibfnamefont {S.}~\bibnamefont
  {Yokota}}, \bibinfo {author} {\bibfnamefont {M.}~\bibnamefont {Adrian}},
  \bibinfo {author} {\bibfnamefont {C.}~\bibnamefont {Aoustin}}, \bibinfo
  {author} {\bibfnamefont {C.}~\bibnamefont {Auletti}}, \bibinfo {author}
  {\bibfnamefont {S.}~\bibnamefont {Aung}}, \bibinfo {author} {\bibfnamefont
  {V.}~\bibnamefont {Bigio}}, \bibinfo {author} {\bibfnamefont
  {N.}~\bibnamefont {Cao}}, \bibinfo {author} {\bibfnamefont {M.}~\bibnamefont
  {Chandler}}, \bibinfo {author} {\bibfnamefont {D.}~\bibnamefont {Chornay}},
  \bibinfo {author} {\bibfnamefont {K.}~\bibnamefont {Christian}}, \bibinfo
  {author} {\bibfnamefont {G.}~\bibnamefont {Clark}}, \bibinfo {author}
  {\bibfnamefont {G.}~\bibnamefont {Collinson}}, \bibinfo {author}
  {\bibfnamefont {T.}~\bibnamefont {Corris}}, \bibinfo {author} {\bibfnamefont
  {A.}~\bibnamefont {De Los Santos}}, \bibinfo {author} {\bibfnamefont
  {R.}~\bibnamefont {Devlin}}, \bibinfo {author} {\bibfnamefont
  {T.}~\bibnamefont {Diaz}}, \bibinfo {author} {\bibfnamefont {T.}~\bibnamefont
  {Dickerson}}, \bibinfo {author} {\bibfnamefont {C.}~\bibnamefont {Dickson}},
  \bibinfo {author} {\bibfnamefont {A.}~\bibnamefont {Diekmann}}, \bibinfo
  {author} {\bibfnamefont {F.}~\bibnamefont {Diggs}}, \bibinfo {author}
  {\bibfnamefont {C.}~\bibnamefont {Duncan}}, \bibinfo {author} {\bibfnamefont
  {A.}~\bibnamefont {Figueroa-Vinas}}, \bibinfo {author} {\bibfnamefont
  {C.}~\bibnamefont {Firman}}, \bibinfo {author} {\bibfnamefont
  {M.}~\bibnamefont {Freeman}}, \bibinfo {author} {\bibfnamefont
  {N.}~\bibnamefont {Galassi}}, \bibinfo {author} {\bibfnamefont
  {K.}~\bibnamefont {Garcia}}, \bibinfo {author} {\bibfnamefont
  {G.}~\bibnamefont {Goodhart}}, \bibinfo {author} {\bibfnamefont
  {D.}~\bibnamefont {Guererro}}, \bibinfo {author} {\bibfnamefont
  {J.}~\bibnamefont {Hageman}}, \bibinfo {author} {\bibfnamefont
  {J.}~\bibnamefont {Hanley}}, \bibinfo {author} {\bibfnamefont
  {E.}~\bibnamefont {Hemminger}}, \bibinfo {author} {\bibfnamefont
  {M.}~\bibnamefont {Holland}}, \bibinfo {author} {\bibfnamefont
  {M.}~\bibnamefont {Hutchins}}, \bibinfo {author} {\bibfnamefont
  {T.}~\bibnamefont {James}}, \bibinfo {author} {\bibfnamefont
  {W.}~\bibnamefont {Jones}}, \bibinfo {author} {\bibfnamefont
  {S.}~\bibnamefont {Kreisler}}, \bibinfo {author} {\bibfnamefont
  {J.}~\bibnamefont {Kujawski}}, \bibinfo {author} {\bibfnamefont
  {V.}~\bibnamefont {Lavu}}, \bibinfo {author} {\bibfnamefont {J.}~\bibnamefont
  {Lobell}}, \bibinfo {author} {\bibfnamefont {E.}~\bibnamefont {LeCompte}},
  \bibinfo {author} {\bibfnamefont {A.}~\bibnamefont {Lukemire}}, \bibinfo
  {author} {\bibfnamefont {E.}~\bibnamefont {MacDonald}}, \bibinfo {author}
  {\bibfnamefont {A.}~\bibnamefont {Mariano}}, \bibinfo {author} {\bibfnamefont
  {T.}~\bibnamefont {Mukai}}, \bibinfo {author} {\bibfnamefont
  {K.}~\bibnamefont {Narayanan}}, \bibinfo {author} {\bibfnamefont
  {Q.}~\bibnamefont {Nguyan}}, \bibinfo {author} {\bibfnamefont
  {M.}~\bibnamefont {Onizuka}}, \bibinfo {author} {\bibfnamefont
  {W.}~\bibnamefont {Paterson}}, \bibinfo {author} {\bibfnamefont
  {S.}~\bibnamefont {Persyn}}, \bibinfo {author} {\bibfnamefont
  {B.}~\bibnamefont {Piepgrass}}, \bibinfo {author} {\bibfnamefont
  {F.}~\bibnamefont {Cheney}}, \bibinfo {author} {\bibfnamefont
  {A.}~\bibnamefont {Rager}}, \bibinfo {author} {\bibfnamefont
  {T.}~\bibnamefont {Raghuram}}, \bibinfo {author} {\bibfnamefont
  {A.}~\bibnamefont {Ramil}}, \bibinfo {author} {\bibfnamefont
  {L.}~\bibnamefont {Reichenthal}}, \bibinfo {author} {\bibfnamefont
  {H.}~\bibnamefont {Rodriguez}}, \bibinfo {author} {\bibfnamefont
  {J.}~\bibnamefont {Rouzaud}}, \bibinfo {author} {\bibfnamefont
  {A.}~\bibnamefont {Rucker}}, \bibinfo {author} {\bibfnamefont
  {Y.}~\bibnamefont {Saito}}, \bibinfo {author} {\bibfnamefont
  {M.}~\bibnamefont {Samara}}, \bibinfo {author} {\bibfnamefont {J.-A.}\
  \bibnamefont {Sauvaud}}, \bibinfo {author} {\bibfnamefont {D.}~\bibnamefont
  {Schuster}}, \bibinfo {author} {\bibfnamefont {M.}~\bibnamefont {Shappirio}},
  \bibinfo {author} {\bibfnamefont {K.}~\bibnamefont {Shelton}}, \bibinfo
  {author} {\bibfnamefont {D.}~\bibnamefont {Sher}}, \bibinfo {author}
  {\bibfnamefont {D.}~\bibnamefont {Smith}}, \bibinfo {author} {\bibfnamefont
  {K.}~\bibnamefont {Smith}}, \bibinfo {author} {\bibfnamefont
  {S.}~\bibnamefont {Smith}}, \bibinfo {author} {\bibfnamefont
  {D.}~\bibnamefont {Steinfeld}}, \bibinfo {author} {\bibfnamefont
  {R.}~\bibnamefont {Szymkiewicz}}, \bibinfo {author} {\bibfnamefont
  {K.}~\bibnamefont {Tanimoto}}, \bibinfo {author} {\bibfnamefont
  {J.}~\bibnamefont {Taylor}}, \bibinfo {author} {\bibfnamefont
  {C.}~\bibnamefont {Tucker}}, \bibinfo {author} {\bibfnamefont
  {K.}~\bibnamefont {Tull}}, \bibinfo {author} {\bibfnamefont {A.}~\bibnamefont
  {Uhl}}, \bibinfo {author} {\bibfnamefont {J.}~\bibnamefont {Vloet}}, \bibinfo
  {author} {\bibfnamefont {P.}~\bibnamefont {Walpole}}, \bibinfo {author}
  {\bibfnamefont {S.}~\bibnamefont {Weidner}}, \bibinfo {author} {\bibfnamefont
  {D.}~\bibnamefont {White}}, \bibinfo {author} {\bibfnamefont
  {G.}~\bibnamefont {Winkert}}, \bibinfo {author} {\bibfnamefont {P.-S.}\
  \bibnamefont {Yeh}},\ and\ \bibinfo {author} {\bibfnamefont {M.}~\bibnamefont
  {Zeuch}},\ }\href {https://doi.org/10.1007/s11214-016-0245-4} {\bibfield
  {journal} {\bibinfo  {journal} {Space Science Reviews}\ }\textbf {\bibinfo
  {volume} {199}},\ \bibinfo {pages} {331} (\bibinfo {year}
  {2016})}\BibitemShut {NoStop}%
\bibitem [{\citenamefont {Gershman}\ \emph {et~al.}(2019)\citenamefont
  {Gershman}, \citenamefont {Dorelli}, \citenamefont {Avanov}, \citenamefont
  {Gliese}, \citenamefont {Barrie}, \citenamefont {Schiff}, \citenamefont
  {Da~Silva}, \citenamefont {Paterson}, \citenamefont {Giles},\ and\
  \citenamefont {Pollock}}]{gershman_systematic_2019}%
  \BibitemOpen
  \bibfield  {author} {\bibinfo {author} {\bibfnamefont {D.~J.}\ \bibnamefont
  {Gershman}}, \bibinfo {author} {\bibfnamefont {J.~C.}\ \bibnamefont
  {Dorelli}}, \bibinfo {author} {\bibfnamefont {L.~A.}\ \bibnamefont {Avanov}},
  \bibinfo {author} {\bibfnamefont {U.}~\bibnamefont {Gliese}}, \bibinfo
  {author} {\bibfnamefont {A.}~\bibnamefont {Barrie}}, \bibinfo {author}
  {\bibfnamefont {C.}~\bibnamefont {Schiff}}, \bibinfo {author} {\bibfnamefont
  {D.~E.}\ \bibnamefont {Da~Silva}}, \bibinfo {author} {\bibfnamefont {W.~R.}\
  \bibnamefont {Paterson}}, \bibinfo {author} {\bibfnamefont {B.~L.}\
  \bibnamefont {Giles}},\ and\ \bibinfo {author} {\bibfnamefont {C.~J.}\
  \bibnamefont {Pollock}},\ }\href {https://doi.org/10.1029/2019JA026980}
  {\bibfield  {journal} {\bibinfo  {journal} {Journal of Geophysical Research:
  Space Physics}\ }\textbf {\bibinfo {volume} {124}},\ \bibinfo {pages} {10345}
  (\bibinfo {year} {2019})}\BibitemShut {NoStop}%
\bibitem [{sup()}]{supplemental_material}%
  \BibitemOpen
  \href@noop {} {}\bibinfo {note} {See Supplemental Material at [URL will be
  inserted by publisher] for additional cuts of the $(\beta_{i\parallel}, R_i)$
  distributions and description of the method to estimate the
  non-Maxwellianity, which includes
  Refs.~\cite{angelopoulos_statistical_1994,richard_are_2022,graham_nonmaxwellianity_2021,pollock_fast_2016}}\BibitemShut
  {NoStop}%
\bibitem [{\citenamefont {Chew}\ \emph {et~al.}(1956)\citenamefont {Chew},
  \citenamefont {Goldberger},\ and\ \citenamefont {Low}}]{chew_boltzmann_1956}%
  \BibitemOpen
  \bibfield  {author} {\bibinfo {author} {\bibfnamefont {G.~F.}\ \bibnamefont
  {Chew}}, \bibinfo {author} {\bibfnamefont {M.~L.}\ \bibnamefont
  {Goldberger}},\ and\ \bibinfo {author} {\bibfnamefont {F.~E.}\ \bibnamefont
  {Low}},\ }\href {https://doi.org/10.1098/rspa.1956.0116} {\bibfield
  {journal} {\bibinfo  {journal} {Proceedings of the Royal Society of London
  Series A}\ }\textbf {\bibinfo {volume} {236}},\ \bibinfo {pages} {112}
  (\bibinfo {year} {1956})}\BibitemShut {NoStop}%
\bibitem [{\citenamefont {Del~Sarto}\ and\ \citenamefont
  {Pegoraro}(2018)}]{del_sarto_shear-induced_2018}%
  \BibitemOpen
  \bibfield  {author} {\bibinfo {author} {\bibfnamefont {D.}~\bibnamefont
  {Del~Sarto}}\ and\ \bibinfo {author} {\bibfnamefont {F.}~\bibnamefont
  {Pegoraro}},\ }\href {https://doi.org/10.1093/mnras/stx3083} {\bibfield
  {journal} {\bibinfo  {journal} {Monthly Notices of the Royal Astronomical
  Society}\ }\textbf {\bibinfo {volume} {475}},\ \bibinfo {pages} {181}
  (\bibinfo {year} {2018})}\BibitemShut {NoStop}%
\bibitem [{\citenamefont {Hellinger}\ \emph {et~al.}(2006)\citenamefont
  {Hellinger}, \citenamefont {Trávníček}, \citenamefont {Kasper},\ and\
  \citenamefont {Lazarus}}]{hellinger_solar_2006}%
  \BibitemOpen
  \bibfield  {author} {\bibinfo {author} {\bibfnamefont {P.}~\bibnamefont
  {Hellinger}}, \bibinfo {author} {\bibfnamefont {P.}~\bibnamefont
  {Trávníček}}, \bibinfo {author} {\bibfnamefont {J.~C.}\ \bibnamefont
  {Kasper}},\ and\ \bibinfo {author} {\bibfnamefont {A.~J.}\ \bibnamefont
  {Lazarus}},\ }\href {https://doi.org/10.1029/2006GL025925} {\bibfield
  {journal} {\bibinfo  {journal} {Geophysical Research Letters}\ }\textbf
  {\bibinfo {volume} {33}},\ \bibinfo {pages} {L09101} (\bibinfo {year}
  {2006})}\BibitemShut {NoStop}%
\bibitem [{\citenamefont {Verscharen}\ \emph {et~al.}(2016)\citenamefont
  {Verscharen}, \citenamefont {Chandran}, \citenamefont {Klein},\ and\
  \citenamefont {Quataert}}]{verscharen_collisionless_2016}%
  \BibitemOpen
  \bibfield  {author} {\bibinfo {author} {\bibfnamefont {D.}~\bibnamefont
  {Verscharen}}, \bibinfo {author} {\bibfnamefont {B.~D.~G.}\ \bibnamefont
  {Chandran}}, \bibinfo {author} {\bibfnamefont {K.~G.}\ \bibnamefont
  {Klein}},\ and\ \bibinfo {author} {\bibfnamefont {E.}~\bibnamefont
  {Quataert}},\ }\href {https://doi.org/10.3847/0004-637X/831/2/128} {\bibfield
   {journal} {\bibinfo  {journal} {The Astrophysical Journal}\ }\textbf
  {\bibinfo {volume} {831}},\ \bibinfo {pages} {128} (\bibinfo {year}
  {2016})}\BibitemShut {NoStop}%
\bibitem [{\citenamefont {Bale}\ \emph {et~al.}(2009)\citenamefont {Bale},
  \citenamefont {Kasper}, \citenamefont {Howes}, \citenamefont {Quataert},
  \citenamefont {Salem},\ and\ \citenamefont {Sundkvist}}]{bale_magnetic_2009}%
  \BibitemOpen
  \bibfield  {author} {\bibinfo {author} {\bibfnamefont {S.~D.}\ \bibnamefont
  {Bale}}, \bibinfo {author} {\bibfnamefont {J.~C.}\ \bibnamefont {Kasper}},
  \bibinfo {author} {\bibfnamefont {G.~G.}\ \bibnamefont {Howes}}, \bibinfo
  {author} {\bibfnamefont {E.}~\bibnamefont {Quataert}}, \bibinfo {author}
  {\bibfnamefont {C.}~\bibnamefont {Salem}},\ and\ \bibinfo {author}
  {\bibfnamefont {D.}~\bibnamefont {Sundkvist}},\ }\href
  {https://doi.org/10.1103/PhysRevLett.103.211101} {\bibfield  {journal}
  {\bibinfo  {journal} {Physical Review Letters}\ }\textbf {\bibinfo {volume}
  {103}},\ \bibinfo {pages} {211101} (\bibinfo {year} {2009})}\BibitemShut
  {NoStop}%
\bibitem [{\citenamefont {Sergeev}\ \emph {et~al.}(2003)\citenamefont
  {Sergeev}, \citenamefont {Runov}, \citenamefont {Baumjohann}, \citenamefont
  {Nakamura}, \citenamefont {Zhang}, \citenamefont {Volwerk}, \citenamefont
  {Balogh}, \citenamefont {Rème}, \citenamefont {Sauvaud}, \citenamefont
  {André},\ and\ \citenamefont {Klecker}}]{sergeev_current_2003}%
  \BibitemOpen
  \bibfield  {author} {\bibinfo {author} {\bibfnamefont {V.~A.}\ \bibnamefont
  {Sergeev}}, \bibinfo {author} {\bibfnamefont {A.}~\bibnamefont {Runov}},
  \bibinfo {author} {\bibfnamefont {W.}~\bibnamefont {Baumjohann}}, \bibinfo
  {author} {\bibfnamefont {R.}~\bibnamefont {Nakamura}}, \bibinfo {author}
  {\bibfnamefont {T.~L.}\ \bibnamefont {Zhang}}, \bibinfo {author}
  {\bibfnamefont {M.}~\bibnamefont {Volwerk}}, \bibinfo {author} {\bibfnamefont
  {A.}~\bibnamefont {Balogh}}, \bibinfo {author} {\bibfnamefont
  {H.}~\bibnamefont {Rème}}, \bibinfo {author} {\bibfnamefont {J.~A.}\
  \bibnamefont {Sauvaud}}, \bibinfo {author} {\bibfnamefont {M.}~\bibnamefont
  {André}},\ and\ \bibinfo {author} {\bibfnamefont {B.}~\bibnamefont
  {Klecker}},\ }\href {https://doi.org/10.1029/2002GL016500} {\bibfield
  {journal} {\bibinfo  {journal} {Geophysical Research Letters}\ }\textbf
  {\bibinfo {volume} {30}},\ \bibinfo {pages} {1327} (\bibinfo {year}
  {2003})}\BibitemShut {NoStop}%
\bibitem [{\citenamefont {Wei}\ \emph {et~al.}(2019)\citenamefont {Wei},
  \citenamefont {Huang}, \citenamefont {Rong}, \citenamefont {Yuan},
  \citenamefont {Jiang}, \citenamefont {Deng}, \citenamefont {Zhou},
  \citenamefont {Fu}, \citenamefont {Yu}, \citenamefont {Xu}, \citenamefont
  {He},\ and\ \citenamefont {Deng}}]{wei_observations_2019}%
  \BibitemOpen
  \bibfield  {author} {\bibinfo {author} {\bibfnamefont {Y.~Y.}\ \bibnamefont
  {Wei}}, \bibinfo {author} {\bibfnamefont {S.~Y.}\ \bibnamefont {Huang}},
  \bibinfo {author} {\bibfnamefont {Z.~J.}\ \bibnamefont {Rong}}, \bibinfo
  {author} {\bibfnamefont {Z.~G.}\ \bibnamefont {Yuan}}, \bibinfo {author}
  {\bibfnamefont {K.}~\bibnamefont {Jiang}}, \bibinfo {author} {\bibfnamefont
  {X.~H.}\ \bibnamefont {Deng}}, \bibinfo {author} {\bibfnamefont
  {M.}~\bibnamefont {Zhou}}, \bibinfo {author} {\bibfnamefont {H.~S.}\
  \bibnamefont {Fu}}, \bibinfo {author} {\bibfnamefont {X.~D.}\ \bibnamefont
  {Yu}}, \bibinfo {author} {\bibfnamefont {S.~B.}\ \bibnamefont {Xu}}, \bibinfo
  {author} {\bibfnamefont {L.~H.}\ \bibnamefont {He}},\ and\ \bibinfo {author}
  {\bibfnamefont {D.}~\bibnamefont {Deng}},\ }\href
  {https://doi.org/10.3847/2041-8213/ab0f28} {\bibfield  {journal} {\bibinfo
  {journal} {The Astrophysical Journal}\ }\textbf {\bibinfo {volume} {874}},\
  \bibinfo {pages} {L18} (\bibinfo {year} {2019})}\BibitemShut {NoStop}%
\bibitem [{\citenamefont {Richard}\ \emph {et~al.}(2021)\citenamefont
  {Richard}, \citenamefont {Khotyaintsev}, \citenamefont {Graham},
  \citenamefont {Sitnov}, \citenamefont {Le~Contel},\ and\ \citenamefont
  {Lindqvist}}]{richard_observations_2021}%
  \BibitemOpen
  \bibfield  {author} {\bibinfo {author} {\bibfnamefont {L.}~\bibnamefont
  {Richard}}, \bibinfo {author} {\bibfnamefont {Y.~V.}\ \bibnamefont
  {Khotyaintsev}}, \bibinfo {author} {\bibfnamefont {D.~B.}\ \bibnamefont
  {Graham}}, \bibinfo {author} {\bibfnamefont {M.~I.}\ \bibnamefont {Sitnov}},
  \bibinfo {author} {\bibfnamefont {O.}~\bibnamefont {Le~Contel}},\ and\
  \bibinfo {author} {\bibfnamefont {P.}~\bibnamefont {Lindqvist}},\ }\href
  {https://doi.org/10.1029/2021JA029152} {\bibfield  {journal} {\bibinfo
  {journal} {Journal of Geophysical Research: Space Physics}\ }\textbf
  {\bibinfo {volume} {126}},\ \bibinfo {pages} {e29152} (\bibinfo {year}
  {2021})}\BibitemShut {NoStop}%
\bibitem [{\citenamefont {Artemyev}\ \emph {et~al.}(2010)\citenamefont
  {Artemyev}, \citenamefont {Petrukovich}, \citenamefont {Nakamura},\ and\
  \citenamefont {Zelenyi}}]{artemyev_proton_2010}%
  \BibitemOpen
  \bibfield  {author} {\bibinfo {author} {\bibfnamefont {A.~V.}\ \bibnamefont
  {Artemyev}}, \bibinfo {author} {\bibfnamefont {A.~A.}\ \bibnamefont
  {Petrukovich}}, \bibinfo {author} {\bibfnamefont {R.}~\bibnamefont
  {Nakamura}},\ and\ \bibinfo {author} {\bibfnamefont {L.~M.}\ \bibnamefont
  {Zelenyi}},\ }\href {https://doi.org/10.1029/2010JA015702} {\bibfield
  {journal} {\bibinfo  {journal} {Journal of Geophysical Research: Space
  Physics}\ }\textbf {\bibinfo {volume} {115}},\ \bibinfo {pages} {A12255}
  (\bibinfo {year} {2010})}\BibitemShut {NoStop}%
\bibitem [{\citenamefont {Graham}\ \emph {et~al.}(2021)\citenamefont {Graham},
  \citenamefont {Khotyaintsev}, \citenamefont {André}, \citenamefont
  {Vaivads}, \citenamefont {Chasapis}, \citenamefont {Matthaeus}, \citenamefont
  {Retinò}, \citenamefont {Valentini},\ and\ \citenamefont
  {Gershman}}]{graham_nonmaxwellianity_2021}%
  \BibitemOpen
  \bibfield  {author} {\bibinfo {author} {\bibfnamefont {D.~B.}\ \bibnamefont
  {Graham}}, \bibinfo {author} {\bibfnamefont {Y.~V.}\ \bibnamefont
  {Khotyaintsev}}, \bibinfo {author} {\bibfnamefont {M.}~\bibnamefont
  {André}}, \bibinfo {author} {\bibfnamefont {A.}~\bibnamefont {Vaivads}},
  \bibinfo {author} {\bibfnamefont {A.}~\bibnamefont {Chasapis}}, \bibinfo
  {author} {\bibfnamefont {W.~H.}\ \bibnamefont {Matthaeus}}, \bibinfo {author}
  {\bibfnamefont {A.}~\bibnamefont {Retinò}}, \bibinfo {author} {\bibfnamefont
  {F.}~\bibnamefont {Valentini}},\ and\ \bibinfo {author} {\bibfnamefont
  {D.~J.}\ \bibnamefont {Gershman}},\ }\href
  {https://doi.org/10.1029/2021JA029260} {\bibfield  {journal} {\bibinfo
  {journal} {Journal of Geophysical Research: Space Physics}\ }\textbf
  {\bibinfo {volume} {126}},\ \bibinfo {pages} {e29260} (\bibinfo {year}
  {2021})}\BibitemShut {NoStop}%
\bibitem [{\citenamefont {Bowling}\ and\ \citenamefont
  {Wolf}(1974)}]{bowling_motion_1974}%
  \BibitemOpen
  \bibfield  {author} {\bibinfo {author} {\bibfnamefont {S.}~\bibnamefont
  {Bowling}}\ and\ \bibinfo {author} {\bibfnamefont {R.}~\bibnamefont {Wolf}},\
  }\href {https://doi.org/10.1016/0032-0633(74)90139-1} {\bibfield  {journal}
  {\bibinfo  {journal} {Planetary and Space Science}\ }\textbf {\bibinfo
  {volume} {22}},\ \bibinfo {pages} {673} (\bibinfo {year} {1974})}\BibitemShut
  {NoStop}%
\bibitem [{\citenamefont {Sergeev}\ \emph {et~al.}(1998)\citenamefont
  {Sergeev}, \citenamefont {Angelopoulos}, \citenamefont {Carlson},\ and\
  \citenamefont {Sutcliffe}}]{sergeev_current_1998}%
  \BibitemOpen
  \bibfield  {author} {\bibinfo {author} {\bibfnamefont {S.}~\bibnamefont
  {Sergeev}, \bibfnamefont {V.~A.}}, \bibinfo {author} {\bibfnamefont
  {V.}~\bibnamefont {Angelopoulos}}, \bibinfo {author} {\bibfnamefont
  {C.}~\bibnamefont {Carlson}},\ and\ \bibinfo {author} {\bibfnamefont
  {P.}~\bibnamefont {Sutcliffe}},\ }\href {https://doi.org/10.1029/97JA02093}
  {\bibfield  {journal} {\bibinfo  {journal} {Journal of Geophysical Research:
  Space Physics}\ }\textbf {\bibinfo {volume} {103}},\ \bibinfo {pages} {9177}
  (\bibinfo {year} {1998})}\BibitemShut {NoStop}%
\bibitem [{\citenamefont {Asano}\ \emph {et~al.}(2003)\citenamefont {Asano},
  \citenamefont {Mukai}, \citenamefont {Hoshino}, \citenamefont {Saito},
  \citenamefont {Hayakawa},\ and\ \citenamefont
  {Nagai}}]{asano_evolution_2003}%
  \BibitemOpen
  \bibfield  {author} {\bibinfo {author} {\bibfnamefont {Y.}~\bibnamefont
  {Asano}}, \bibinfo {author} {\bibfnamefont {T.}~\bibnamefont {Mukai}},
  \bibinfo {author} {\bibfnamefont {M.}~\bibnamefont {Hoshino}}, \bibinfo
  {author} {\bibfnamefont {Y.}~\bibnamefont {Saito}}, \bibinfo {author}
  {\bibfnamefont {H.}~\bibnamefont {Hayakawa}},\ and\ \bibinfo {author}
  {\bibfnamefont {T.}~\bibnamefont {Nagai}},\ }\bibfield  {journal} {\bibinfo
  {journal} {Journal of Geophysical Research: Space Physics}\ }\textbf
  {\bibinfo {volume} {108}},\ \href {https://doi.org/10.1029/2002JA009785}
  {10.1029/2002JA009785} (\bibinfo {year} {2003})\BibitemShut {NoStop}%
\bibitem [{\citenamefont {Vörös}\ \emph {et~al.}(2006)\citenamefont
  {Vörös}, \citenamefont {Baumjohann}, \citenamefont {Nakamura},
  \citenamefont {Volwerk},\ and\ \citenamefont {Runov}}]{voros_bursty_2006}%
  \BibitemOpen
  \bibfield  {author} {\bibinfo {author} {\bibfnamefont {Z.}~\bibnamefont
  {Vörös}}, \bibinfo {author} {\bibfnamefont {W.}~\bibnamefont {Baumjohann}},
  \bibinfo {author} {\bibfnamefont {R.}~\bibnamefont {Nakamura}}, \bibinfo
  {author} {\bibfnamefont {M.}~\bibnamefont {Volwerk}},\ and\ \bibinfo {author}
  {\bibfnamefont {A.}~\bibnamefont {Runov}},\ }\href
  {https://doi.org/10.1007/s11214-006-6987-7} {\bibfield  {journal} {\bibinfo
  {journal} {Space Science Reviews}\ }\textbf {\bibinfo {volume} {122}},\
  \bibinfo {pages} {301} (\bibinfo {year} {2006})}\BibitemShut {NoStop}%
\bibitem [{\citenamefont {Maruca}\ \emph {et~al.}(2018)\citenamefont {Maruca},
  \citenamefont {Chasapis}, \citenamefont {Gary}, \citenamefont
  {Bandyopadhyay}, \citenamefont {Chhiber}, \citenamefont {Parashar},
  \citenamefont {Matthaeus}, \citenamefont {Shay}, \citenamefont {Burch},
  \citenamefont {Moore}, \citenamefont {Pollock}, \citenamefont {Giles},
  \citenamefont {Paterson}, \citenamefont {Dorelli}, \citenamefont {Gershman},
  \citenamefont {Torbert}, \citenamefont {Russell},\ and\ \citenamefont
  {Strangeway}}]{maruca_mms_2018}%
  \BibitemOpen
  \bibfield  {author} {\bibinfo {author} {\bibfnamefont {B.~A.}\ \bibnamefont
  {Maruca}}, \bibinfo {author} {\bibfnamefont {A.}~\bibnamefont {Chasapis}},
  \bibinfo {author} {\bibfnamefont {S.~P.}\ \bibnamefont {Gary}}, \bibinfo
  {author} {\bibfnamefont {R.}~\bibnamefont {Bandyopadhyay}}, \bibinfo {author}
  {\bibfnamefont {R.}~\bibnamefont {Chhiber}}, \bibinfo {author} {\bibfnamefont
  {T.~N.}\ \bibnamefont {Parashar}}, \bibinfo {author} {\bibfnamefont {W.~H.}\
  \bibnamefont {Matthaeus}}, \bibinfo {author} {\bibfnamefont {M.~A.}\
  \bibnamefont {Shay}}, \bibinfo {author} {\bibfnamefont {J.~L.}\ \bibnamefont
  {Burch}}, \bibinfo {author} {\bibfnamefont {T.~E.}\ \bibnamefont {Moore}},
  \bibinfo {author} {\bibfnamefont {C.~J.}\ \bibnamefont {Pollock}}, \bibinfo
  {author} {\bibfnamefont {B.~J.}\ \bibnamefont {Giles}}, \bibinfo {author}
  {\bibfnamefont {W.~R.}\ \bibnamefont {Paterson}}, \bibinfo {author}
  {\bibfnamefont {J.}~\bibnamefont {Dorelli}}, \bibinfo {author} {\bibfnamefont
  {D.~J.}\ \bibnamefont {Gershman}}, \bibinfo {author} {\bibfnamefont {R.~B.}\
  \bibnamefont {Torbert}}, \bibinfo {author} {\bibfnamefont {C.~T.}\
  \bibnamefont {Russell}},\ and\ \bibinfo {author} {\bibfnamefont {R.~J.}\
  \bibnamefont {Strangeway}},\ }\href
  {https://doi.org/10.3847/1538-4357/aaddfb} {\bibfield  {journal} {\bibinfo
  {journal} {The Astrophysical Journal}\ }\textbf {\bibinfo {volume} {866}},\
  \bibinfo {pages} {25} (\bibinfo {year} {2018})}\BibitemShut {NoStop}%
\bibitem [{\citenamefont {Kennel}\ and\ \citenamefont
  {Petschek}(1966)}]{kennel_limit_1966}%
  \BibitemOpen
  \bibfield  {author} {\bibinfo {author} {\bibfnamefont {C.~F.}\ \bibnamefont
  {Kennel}}\ and\ \bibinfo {author} {\bibfnamefont {H.~E.}\ \bibnamefont
  {Petschek}},\ }\href {https://doi.org/10.1029/JZ071i001p00001} {\bibfield
  {journal} {\bibinfo  {journal} {Journal of Geophysical Research}\ }\textbf
  {\bibinfo {volume} {71}},\ \bibinfo {pages} {1} (\bibinfo {year}
  {1966})}\BibitemShut {NoStop}%
\bibitem [{\citenamefont {Lichtenberg}\ and\ \citenamefont
  {Lieberman}(1983)}]{lichtenberg_regular_1983}%
  \BibitemOpen
  \bibfield  {author} {\bibinfo {author} {\bibfnamefont {A.~J.}\ \bibnamefont
  {Lichtenberg}}\ and\ \bibinfo {author} {\bibfnamefont {M.~A.}\ \bibnamefont
  {Lieberman}},\ }\href {https://doi.org/10.1007/978-1-4757-4257-2} {\emph
  {\bibinfo {title} {Regular and {Stochastic} {Motion}}}},\ edited by\ \bibinfo
  {editor} {\bibfnamefont {F.}~\bibnamefont {John}}, \bibinfo {editor}
  {\bibfnamefont {J.~E.}\ \bibnamefont {Marsden}},\ and\ \bibinfo {editor}
  {\bibfnamefont {L.}~\bibnamefont {Sirovich}},\ \bibinfo {series} {Applied
  {Mathematical} {Sciences}}, Vol.~\bibinfo {volume} {38}\ (\bibinfo
  {publisher} {Springer New York},\ \bibinfo {year} {1983})\BibitemShut
  {NoStop}%
\bibitem [{\citenamefont {Chen}(1992)}]{chen_nonlinear_1992}%
  \BibitemOpen
  \bibfield  {author} {\bibinfo {author} {\bibfnamefont {J.}~\bibnamefont
  {Chen}},\ }\href {https://doi.org/10.1029/92JA00955} {\bibfield  {journal}
  {\bibinfo  {journal} {Journal of Geophysical Research}\ }\textbf {\bibinfo
  {volume} {97}},\ \bibinfo {pages} {15011} (\bibinfo {year}
  {1992})}\BibitemShut {NoStop}%
\bibitem [{\citenamefont {Zelenyi}\ \emph {et~al.}(2013)\citenamefont
  {Zelenyi}, \citenamefont {Neishtadt}, \citenamefont {Artemyev}, \citenamefont
  {Vainchtein},\ and\ \citenamefont {Malova}}]{zelenyi_quasiadiabatic_2013}%
  \BibitemOpen
  \bibfield  {author} {\bibinfo {author} {\bibfnamefont {L.~M.}\ \bibnamefont
  {Zelenyi}}, \bibinfo {author} {\bibfnamefont {A.~I.}\ \bibnamefont
  {Neishtadt}}, \bibinfo {author} {\bibfnamefont {A.~V.}\ \bibnamefont
  {Artemyev}}, \bibinfo {author} {\bibfnamefont {D.~L.}\ \bibnamefont
  {Vainchtein}},\ and\ \bibinfo {author} {\bibfnamefont {H.~V.}\ \bibnamefont
  {Malova}},\ }\href {https://doi.org/10.3367/UFNe.0183.201304b.0365}
  {\bibfield  {journal} {\bibinfo  {journal} {Physics Uspekhi}\ }\textbf
  {\bibinfo {volume} {56}},\ \bibinfo {pages} {347} (\bibinfo {year}
  {2013})}\BibitemShut {NoStop}%
\bibitem [{\citenamefont {Runov}\ \emph {et~al.}(2005)\citenamefont {Runov},
  \citenamefont {Sergeev}, \citenamefont {Baumjohann}, \citenamefont
  {Nakamura}, \citenamefont {Apatenkov}, \citenamefont {Asano}, \citenamefont
  {Volwerk}, \citenamefont {Vörös}, \citenamefont {Zhang}, \citenamefont
  {Petrukovich}, \citenamefont {Balogh}, \citenamefont {Sauvaud}, \citenamefont
  {Klecker},\ and\ \citenamefont {Rème}}]{runov_electric_2005}%
  \BibitemOpen
  \bibfield  {author} {\bibinfo {author} {\bibfnamefont {A.}~\bibnamefont
  {Runov}}, \bibinfo {author} {\bibfnamefont {V.~A.}\ \bibnamefont {Sergeev}},
  \bibinfo {author} {\bibfnamefont {W.}~\bibnamefont {Baumjohann}}, \bibinfo
  {author} {\bibfnamefont {R.}~\bibnamefont {Nakamura}}, \bibinfo {author}
  {\bibfnamefont {S.}~\bibnamefont {Apatenkov}}, \bibinfo {author}
  {\bibfnamefont {Y.}~\bibnamefont {Asano}}, \bibinfo {author} {\bibfnamefont
  {M.}~\bibnamefont {Volwerk}}, \bibinfo {author} {\bibfnamefont
  {Z.}~\bibnamefont {Vörös}}, \bibinfo {author} {\bibfnamefont {T.~L.}\
  \bibnamefont {Zhang}}, \bibinfo {author} {\bibfnamefont {A.}~\bibnamefont
  {Petrukovich}}, \bibinfo {author} {\bibfnamefont {A.}~\bibnamefont {Balogh}},
  \bibinfo {author} {\bibfnamefont {J.-A.}\ \bibnamefont {Sauvaud}}, \bibinfo
  {author} {\bibfnamefont {B.}~\bibnamefont {Klecker}},\ and\ \bibinfo {author}
  {\bibfnamefont {H.}~\bibnamefont {Rème}},\ }\href
  {https://doi.org/10.5194/angeo-23-1391-2005} {\bibfield  {journal} {\bibinfo
  {journal} {Annales Geophysicae}\ }\textbf {\bibinfo {volume} {23}},\ \bibinfo
  {pages} {1391} (\bibinfo {year} {2005})}\BibitemShut {NoStop}%
\bibitem [{\citenamefont {Chanteur}(1998)}]{chanteur_spatial_1998}%
  \BibitemOpen
  \bibfield  {author} {\bibinfo {author} {\bibfnamefont {G.}~\bibnamefont
  {Chanteur}},\ }\href@noop {} {\bibfield  {journal} {\bibinfo  {journal} {ISSI
  Scientific Reports}\ ,\ \bibinfo {pages} {349}} (\bibinfo {year}
  {1998})}\BibitemShut {NoStop}%
\bibitem [{\citenamefont {Chanteur}\ and\ \citenamefont
  {Harvey}(1998)}]{chanteur_spatial_1998-1}%
  \BibitemOpen
  \bibfield  {author} {\bibinfo {author} {\bibfnamefont {G.}~\bibnamefont
  {Chanteur}}\ and\ \bibinfo {author} {\bibfnamefont {C.~C.}\ \bibnamefont
  {Harvey}},\ }\href@noop {} {\bibfield  {journal} {\bibinfo  {journal} {ISSI
  Scientific Reports}\ ,\ \bibinfo {pages} {371}} (\bibinfo {year}
  {1998})}\BibitemShut {NoStop}%
\bibitem [{\citenamefont {Lukin}\ \emph {et~al.}(2022)\citenamefont {Lukin},
  \citenamefont {Artemyev}, \citenamefont {Vainchtein},\ and\ \citenamefont
  {Petrukovich}}]{lukin_regimes_2022}%
  \BibitemOpen
  \bibfield  {author} {\bibinfo {author} {\bibfnamefont {A.~S.}\ \bibnamefont
  {Lukin}}, \bibinfo {author} {\bibfnamefont {A.~V.}\ \bibnamefont {Artemyev}},
  \bibinfo {author} {\bibfnamefont {D.~L.}\ \bibnamefont {Vainchtein}},\ and\
  \bibinfo {author} {\bibfnamefont {A.~A.}\ \bibnamefont {Petrukovich}},\
  }\href {https://doi.org/10.1103/PhysRevE.106.065205} {\bibfield  {journal}
  {\bibinfo  {journal} {Physical Review E}\ }\textbf {\bibinfo {volume}
  {106}},\ \bibinfo {pages} {065205} (\bibinfo {year} {2022})}\BibitemShut
  {NoStop}%
\bibitem [{\citenamefont {Schindler}(1965)}]{schindler_adiabatic_1965}%
  \BibitemOpen
  \bibfield  {author} {\bibinfo {author} {\bibfnamefont {K.}~\bibnamefont
  {Schindler}},\ }\href {https://doi.org/10.1063/1.1704282} {\bibfield
  {journal} {\bibinfo  {journal} {Journal of Mathematical Physics}\ }\textbf
  {\bibinfo {volume} {6}},\ \bibinfo {pages} {313} (\bibinfo {year}
  {1965})}\BibitemShut {NoStop}%
\bibitem [{\citenamefont {Sonnerup}(1971)}]{sonnerup_adiabatic_1971}%
  \BibitemOpen
  \bibfield  {author} {\bibinfo {author} {\bibfnamefont {B.~U.~O.}\
  \bibnamefont {Sonnerup}},\ }\href {https://doi.org/10.1029/JA076i034p08211}
  {\bibfield  {journal} {\bibinfo  {journal} {Journal of Geophysical Research}\
  }\textbf {\bibinfo {volume} {76}},\ \bibinfo {pages} {8211} (\bibinfo {year}
  {1971})}\BibitemShut {NoStop}%
\bibitem [{\citenamefont {Artemyev}\ \emph {et~al.}(2016)\citenamefont
  {Artemyev}, \citenamefont {Angelopoulos},\ and\ \citenamefont
  {Runov}}]{artemyev_radial_2016}%
  \BibitemOpen
  \bibfield  {author} {\bibinfo {author} {\bibfnamefont {A.~V.}\ \bibnamefont
  {Artemyev}}, \bibinfo {author} {\bibfnamefont {V.}~\bibnamefont
  {Angelopoulos}},\ and\ \bibinfo {author} {\bibfnamefont {A.}~\bibnamefont
  {Runov}},\ }\href {https://doi.org/10.1002/2016JA022480} {\bibfield
  {journal} {\bibinfo  {journal} {Journal of Geophysical Research: Space
  Physics}\ }\textbf {\bibinfo {volume} {121}},\ \bibinfo {pages} {4017}
  (\bibinfo {year} {2016})}\BibitemShut {NoStop}%
\bibitem [{\citenamefont {Aunai}\ \emph {et~al.}(2011)\citenamefont {Aunai},
  \citenamefont {Belmont},\ and\ \citenamefont {Smets}}]{aunai_proton_2011}%
  \BibitemOpen
  \bibfield  {author} {\bibinfo {author} {\bibfnamefont {N.}~\bibnamefont
  {Aunai}}, \bibinfo {author} {\bibfnamefont {G.}~\bibnamefont {Belmont}},\
  and\ \bibinfo {author} {\bibfnamefont {R.}~\bibnamefont {Smets}},\ }\href
  {https://doi.org/10.1029/2011JA016688} {\bibfield  {journal} {\bibinfo
  {journal} {Journal of Geophysical Research: Space Physics}\ }\textbf
  {\bibinfo {volume} {116}},\ \bibinfo {pages} {A09232} (\bibinfo {year}
  {2011})}\BibitemShut {NoStop}%
\bibitem [{\citenamefont {Divin}\ \emph {et~al.}(2016)\citenamefont {Divin},
  \citenamefont {Khotyaintsev}, \citenamefont {Vaivads}, \citenamefont
  {André}, \citenamefont {Toledo-Redondo}, \citenamefont {Markidis},\ and\
  \citenamefont {Lapenta}}]{divin_three-scale_2016}%
  \BibitemOpen
  \bibfield  {author} {\bibinfo {author} {\bibfnamefont {A.}~\bibnamefont
  {Divin}}, \bibinfo {author} {\bibfnamefont {Y.~V.}\ \bibnamefont
  {Khotyaintsev}}, \bibinfo {author} {\bibfnamefont {A.}~\bibnamefont
  {Vaivads}}, \bibinfo {author} {\bibfnamefont {M.}~\bibnamefont {André}},
  \bibinfo {author} {\bibfnamefont {S.}~\bibnamefont {Toledo-Redondo}},
  \bibinfo {author} {\bibfnamefont {S.}~\bibnamefont {Markidis}},\ and\
  \bibinfo {author} {\bibfnamefont {G.}~\bibnamefont {Lapenta}},\ }\href
  {https://doi.org/10.1002/2016JA023606} {\bibfield  {journal} {\bibinfo
  {journal} {Journal of Geophysical Research: Space Physics}\ }\textbf
  {\bibinfo {volume} {121}},\ \bibinfo {pages} {12,001} (\bibinfo {year}
  {2016})}\BibitemShut {NoStop}%
\bibitem [{\citenamefont {Artemyev}\ \emph {et~al.}(2014)\citenamefont
  {Artemyev}, \citenamefont {Neishtadt},\ and\ \citenamefont
  {Zelenyi}}]{artemyev_rapid_2014}%
  \BibitemOpen
  \bibfield  {author} {\bibinfo {author} {\bibfnamefont {A.~V.}\ \bibnamefont
  {Artemyev}}, \bibinfo {author} {\bibfnamefont {A.~I.}\ \bibnamefont
  {Neishtadt}},\ and\ \bibinfo {author} {\bibfnamefont {L.~M.}\ \bibnamefont
  {Zelenyi}},\ }\href {https://doi.org/10.1103/PhysRevE.89.060902} {\bibfield
  {journal} {\bibinfo  {journal} {Physical Review E}\ }\textbf {\bibinfo
  {volume} {89}},\ \bibinfo {pages} {060902} (\bibinfo {year}
  {2014})}\BibitemShut {NoStop}%
\bibitem [{\citenamefont {Artemyev}\ \emph {et~al.}(2020)\citenamefont
  {Artemyev}, \citenamefont {Neishtadt}, \citenamefont {Vasiliev},
  \citenamefont {Angelopoulos}, \citenamefont {Vinogradov},\ and\ \citenamefont
  {Zelenyi}}]{artemyev_superfast_2020}%
  \BibitemOpen
  \bibfield  {author} {\bibinfo {author} {\bibfnamefont {A.~V.}\ \bibnamefont
  {Artemyev}}, \bibinfo {author} {\bibfnamefont {A.~I.}\ \bibnamefont
  {Neishtadt}}, \bibinfo {author} {\bibfnamefont {A.~A.}\ \bibnamefont
  {Vasiliev}}, \bibinfo {author} {\bibfnamefont {V.}~\bibnamefont
  {Angelopoulos}}, \bibinfo {author} {\bibfnamefont {A.~A.}\ \bibnamefont
  {Vinogradov}},\ and\ \bibinfo {author} {\bibfnamefont {L.~M.}\ \bibnamefont
  {Zelenyi}},\ }\href {https://doi.org/10.1103/PhysRevE.102.033201} {\bibfield
  {journal} {\bibinfo  {journal} {Physical Review E}\ }\textbf {\bibinfo
  {volume} {102}},\ \bibinfo {pages} {033201} (\bibinfo {year}
  {2020})}\BibitemShut {NoStop}%
\bibitem [{\citenamefont {Reale}(2014)}]{reale_coronal_2014}%
  \BibitemOpen
  \bibfield  {author} {\bibinfo {author} {\bibfnamefont {F.}~\bibnamefont
  {Reale}},\ }\href {https://doi.org/10.12942/lrsp-2014-4} {\bibfield
  {journal} {\bibinfo  {journal} {Living Reviews in Solar Physics}\ }\textbf
  {\bibinfo {volume} {11}},\ \bibinfo {pages} {4} (\bibinfo {year}
  {2014})}\BibitemShut {NoStop}%
\bibitem [{Note1()}]{Note1}%
  \BibitemOpen
  \bibinfo {note} {See \protect \url
  {https://lasp.colorado.edu/mms/sdc/public}.}\BibitemShut {Stop}%
\bibitem [{Note2()}]{Note2}%
  \BibitemOpen
  \bibinfo {note} {See \protect \url
  {https://pypi.org/project/pyrfu/}.}\BibitemShut {Stop}%
\end{thebibliography}%

\end{document}